\documentclass{aa}  
\usepackage{graphicx}
\usepackage{txfonts}
\usepackage{hyperref}
\usepackage{makecell}
\usepackage{subfiles}
\usepackage{siunitx}
\usepackage[normalem]{ulem}
\usepackage{threeparttable}
\usepackage{booktabs}
\usepackage{multirow}
\DeclareMathAlphabet{\mathcal}{OMS}{cmsy}{m}{n}
\newcommand\teff{\ensuremath{T_{\textup{eff}}}\xspace}
\newcommand\logg{\ensuremath{\log(g)}\xspace}
\newcommand\muHz{\ensuremath{\mu \mathrm{Hz}}\xspace}
\newcommand\numax{\ensuremath{\nu_{\textup{max}}}\xspace}
\newcommand\dnu{\ensuremath{\Delta\nu}\xspace}

\newcommand\FeH{\ensuremath{[\mathrm{Fe}/\mathrm{H}]}\xspace}

\newcommand\kepler{\textit{Kepler}\xspace}

\usepackage{xcolor}
\usepackage{listings}
\lstdefinelanguage{Stan}{
  morekeywords={
    data, parameters, model, transformed, generated,
    int, real, vector, matrix, array,
    if, else, for, in
  },
  sensitive=true,
  morecomment=[l]{//},
  morecomment=[s]{/*}{*/},
  morestring=[b]",
}

\begin{document} 
    \title{Granulation signatures as seen by \kepler short-cadence data}
    \subtitle{II. A hierarchical route to inferring stellar radii from granulation}
    
   \author{J. R. Larsen\inst{1}\fnmsep\thanks{E-mail: jensrl@phys.au.dk}
            \and
            M. S. Lundkvist\inst{1}
            \and
            G. R. Davies\inst{2}
            \and
            M. B. Nielsen\inst{2}
        }
        
   \institute{Stellar Astrophysics Centre (SAC), Department of Physics and Astronomy, Aarhus University, Ny Munkegade 120, 8000 Aarhus C, DK 
              \and
              School of Physics and Astronomy, University of Birmingham, Edgbaston B15 2TT, UK 
            }
   \date{Received 27 March, 2026; Accepted 25 May, 2026}

  \abstract
    {Stellar granulation arises from near-surface convection and is imprinted in stellar photometric time series. Yet the links between granulation observables and fundamental stellar properties remain underexploited.
   }
   {We aim to establish a statistically robust framework for inferring stellar radii directly from granulation signals in long-duration space-based photometry with aid from stellar atmospheric parameters.
   }
   {We construct a Bayesian hierarchical model to connect stellar radius and granulation, relating radius through regression to the total granulation amplitude, primary characteristic frequency of the granulation, stellar effective temperature, and surface metallicity. The derivation is performed separately for three granulation models, propagating the complete marginal posteriors of the granulation parameters to account for intrinsic dispersion of the derived relations. Each background model yields a unique radius posterior, subsequently combined using Bayesian evidences as weights. This produces radius posteriors that best represent the given star by marginalising over the different background models. 
   }
   {The granulation-radius relations were derived from a heterogeneous sample of 363 stars, combining seismic and interferometric targets from multiple sources with their associated systematics. Application to an independent sample of 367 stars recovers the reference radii within $1\sigma$ in ${\approx}73\%$ of cases. The distribution of residuals is consistent with a well-calibrated and unbiased inference. Across applications, the granulation-inferred radii achieve a precision of ${\approx}10\%$. The agreement with seismic and interferometric benchmarks demonstrates that granulation carries predictive information on stellar radii at a level comparable to several established techniques.
   }
  {This work enables the inference of stellar radii from granulation signals; directly applicable to data from \kepler, TESS, and the upcoming ESA PLATO mission. Using granulation as a structural diagnostic enables radius estimation for stars which may otherwise be challenging to characterise, thereby providing a complementary approach to stellar characterisation across diverse populations.
  }
   \titlerunning{A hierarchical route to inferring stellar radii from granulation}  
   \maketitle


\section{Introduction}\label{sec:Intro}
Stellar radii are among the fundamental physical parameters in astrophysics. They enter directly in the determination of luminosities, surface gravities, and effective temperatures, and they set the physical scale for stellar structure and evolution models \citep{Kippenhahn13}. Moreover, precise stellar radii are crucial for a wide range of applications, from calibrating stellar evolution theory \citep{Torres10,Bellinger19} to characterizing exoplanets, where planetary sizes and bulk densities scale directly with the properties of their host stars \citep{Stassun17,Adibekyan18}. Improving the availability and reliability of stellar radii therefore has broad implications across stellar and exoplanetary astrophysics.

Asteroseismology has emerged as one of the most powerful tools for determining stellar radii. The detection and characterisation of solar-like oscillations provide direct constraints on the properties of stars. Often these are obtained through the use of scaling relations involving the global asteroseismic properties known as the frequency of maximum oscillation power, \numax, and the large frequency separation, \dnu (see e.g. \citealt{Chaplin13} for a review). Asteroseismic signals have thus enabled precise radii to be determined for large samples of stars observed by missions such as CoRoT \citep{Baglin06}, \kepler \citep{Borucki10}, and the Transiting Exoplanet Survey Satellite (TESS; \citealt{Ricker14}). However, the applicability of asteroseismology is ultimately limited by detectability (see e.g. \citealt{Chaplin11b}). Oscillation signals are not always measurable, either because of insufficient signal-to-noise (S/N), short observing baselines, or because the stars themselves do not exhibit detectable oscillations in the available data. This motivates the exploration of alternative variability diagnostics that can be measured in a wider range of stars.

One such signal is stellar granulation, the stochastic brightness variations produced by the convective motions at the stellar surface \citep{Nordlund09,Magic14_unpublished}. Granulation leaves a characteristic imprint in the frequency domain, typically described by one or more background components in stellar power density spectra (PDSs). The presence of oscillations, and their reliability for inferencing stellar radii, is sensitive to the level of photon noise \citep{Chaplin11b}. Granulation, by contrast, produces correlated variability in the time series spread over a wide range in frequency \citep{Harvey85,Mathur11,Kallinger2014}. This makes its properties easier to measure reliably, less sensitive to the photon noise, and is in principle applicable to all stars with convective envelopes, thereby establishing granulation as a promising diagnostic of stellar structure. Previous studies have demonstrated that granulation-based metrics can be used to infer surface gravities \citep{Bastien16,Bugnet18,Pande18}. These approaches highlight the potential of granulation as an alternative or complementary probe of stellar parameters.

In \citet{Larsen26} (hereafter Paper~I), we presented an analysis of granulation signatures across a sample of 753 stars observed by \kepler in short cadence. The studied sample originated from \citet{Sayeed25}, which additionally compiles available atmospheric, seismic, and determined stellar parameters. Paper~I performed detailed granulation background inferences using three different model prescriptions within a Bayesian framework. The analysis yielded posterior probability distributions and Bayesian evidences for each star and model combination, providing a statistically consistent characterisation of granulation properties across the sample. In the present work, we build upon this foundation and explore whether granulation signals can be used to infer stellar radii. We seek to utilise hierarchical modelling to derive scaling relations for predicting stellar radii directly from the signals of granulation, in combination with stellar effective temperatures and surface metallicities. This approach enables a self-consistent derivation of population-level granulation-radius relations, carefully propagating uncertainties and correlations from individual stars, thereby providing a novel, independent, and widely applicable method for estimating stellar radii.

The primary scope of this paper is therefore methodological: we introduce and develop a hierarchical framework for deriving stellar radii from granulation signals. We demonstrate its performance using the sample of solar-like oscillators from Paper~I, supplemented with interferometric radii of a sample of stars observed by TESS. We focus on establishing the statistical foundations of the method, assessing its internal consistency, and identifying its current limitations. We furthermore discuss the benefits of potential future applicability to broader stellar populations and observational datasets, especially in connection with the upcoming PLAnetary Transits and Oscillations of stars mission (PLATO; \citealt{Rauer25}).

This paper is structured as follows. In Sect.\ref{sec:CalibValid}, we briefly motivate our approach and the relevant findings from Paper~I, before presenting the available data and defining two subsamples for deriving and applying the methodology. The hierarchical methodology is outlined throughout Sect.~\ref{sec:Methods} and afterwards used to derive the connections between stellar radii and granulation signals in Sect.~\ref{sec:Derivation}. In Sect.~\ref{sec:Application} we validate the derived methodology through extensive application and statistical diagnostics. Lastly, in Sect.~\ref{sec:Discuss} we discuss the applicability and future use-cases of the derived methodology, before concluding in Sect.~\ref{sec:Conclusion}.

\section{Calibration and validation sets}\label{sec:CalibValid}
\begin{figure*}[t]
    \includegraphics[width=17cm]{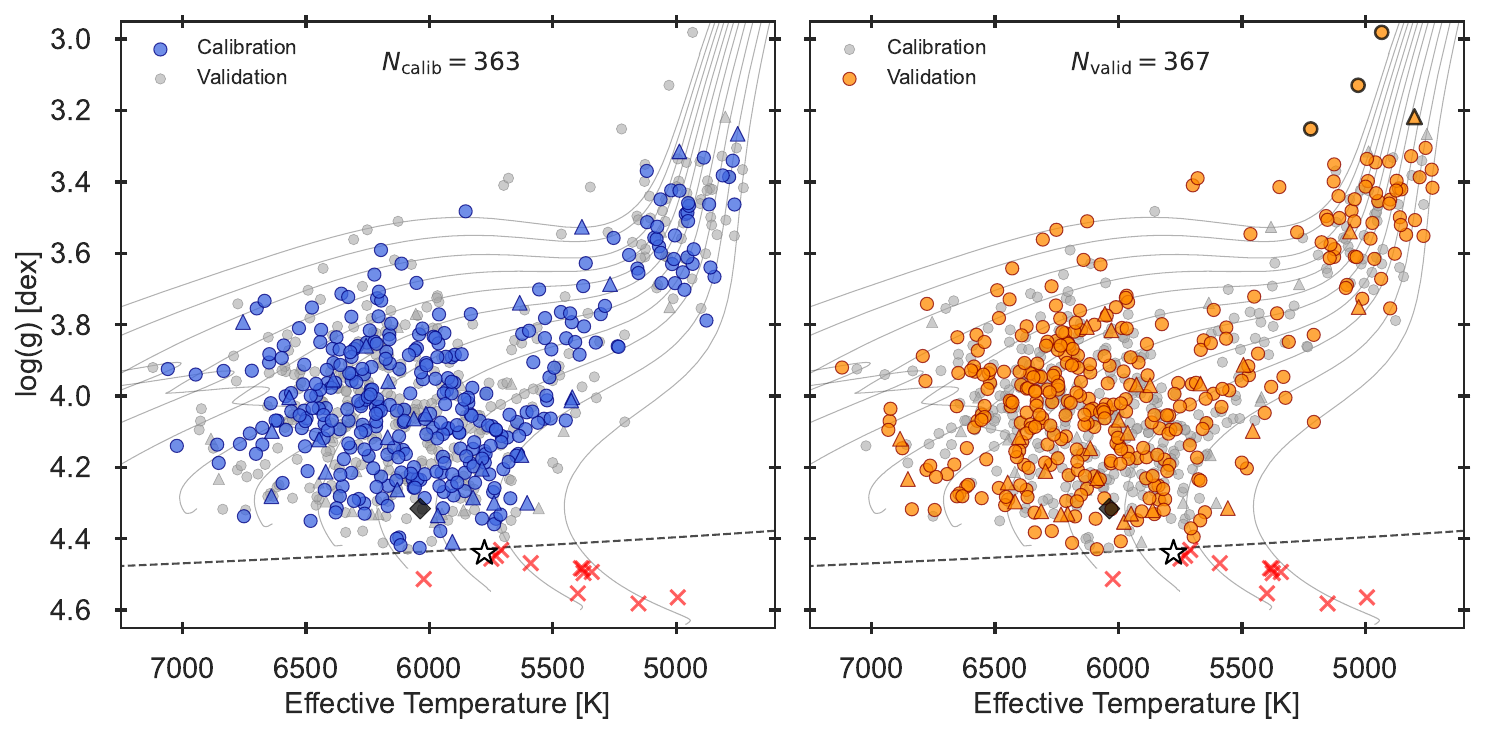}
    \centering
    \vspace*{-1.5mm}
    \caption{Kiel diagrams of the studied sample, each panel highlighting the calibration (left) and validation (right) samples. The stars with asteroseismic and interferometric radii are plotted with circle and triangle symbols, respectively. The $\numax < 3000$ \muHz criteria is indicated by the dashed black lines, and the stars from Paper~I which were removed as a result are shown as the red crosses. The Sun and reference star KIC~6106415 are shown as the white star and black diamond symbols, respectively. The four RGB stars forced into the validation sample are highlighted with black outlines in the right-hand panel.}
    \label{fig:KielOverview}
    \vspace*{-2mm}
\end{figure*}
In this work, our objective was to develop a broadly applicable method for inferring stellar radii from granulation signals. For this purpose we required one sample to train and derive our methodology and another for application tests, referred to as the calibration and validation samples, respectively. 

Stellar properties in our sample are derived using a range of methodologies: radii may come from detailed asteroseismic modelling or scaling-relation approaches, while effective temperatures are provided by multiple surveys and pipelines tied to different temperature scales. These methodological differences, together with variations in data quality, introduce distinct and non-negligible systematics. To ensure that our methodology is constructed and tested on samples reflecting realistic use cases, we therefore designed the calibration and validation strategy to preserve this heterogeneity. This was done using random selection across the different source classes, such that both subsets retain the diversity of the underlying systematics of the methods with which they were measured. 

Moreover, in Paper~I we showed that a decoupling between the granulation and oscillation timescales occur for main-sequence (MS) stars cooler than the Sun, i.e. from around $\numax \gtrsim 3000$ \muHz. The information carried by the granulation signals thus become less constraining for the fundamental stellar parameters. As a result we chose to only consider stars with $\numax < 3000$ \muHz. 

One star was removed prior to constructing the samples: KIC~6106415 (Perky). This star is instead adopted as the reference star for the methodology presented later (see Sect.~\ref{subsec:PerkyRelSpace} for details). It exhibits strong solar-like oscillations, low activity, a representative value of $\numax = 2248.6\pm4.6$ \muHz, and is among the best-characterised stars in the Legacy sample \citep{Lund2017}. Additionally, Perky also has an interferometric radius estimate from \citet{Soubiran22} which we later compare to (see Table~\ref{tab:Inter3adds}).

\subsection{Stars with seismic radii}\label{subsec:seissample}
The sample in Paper~I was created from the catalogue of \citet{Sayeed25}, and we thereby have both atmospheric measurements and asteroseismically determined fundamental parameters available for a large portion of the sample. The seismic sample consists of stars from various sources: the Legacy sample \citep{Lund2017}, the KAGES sample \citep{Aguirre15,Davies2016}, the APOKASC~1 sample \citep{Serenelli17}, and other alternative sources as presented in \citet{Sayeed25}. The stars thus range from the most well-characterised solar-like oscillators on the MS and sub-giant branch (SGB) to those with weakly detected oscillations signals providing only global asteroseismic parameters. 

We require the stars used in this work to have an asteroseismic estimate of the stellar radius and an effective temperature measurement available; if a star does not it was removed. This left us with 677 stars remaining from Paper~I with asteroseismic radii and effective temperature estimates, accompanied by full posterior probability distributions for all granulation parameters and Bayesian evidences across three different granulation background models.

\subsection{Stars with interferometric radii}
As motivated above, we want the developed methodology to account for the systematics of heterogeneous data. To this end, we constructed a sample of stars with interferometric radii (see Appendix~\ref{app:intersamp}). This enables both the derivation of the methodology and its validation against an independent radius determination, characterised by distinctly different assumptions and systematics compared to the seismic sample.

The majority of the interferometric sample is made up of 54 TESS stars. To allow for their inclusion we had to repeat the process of Paper~I to obtain the granulation posteriors. We applied the developed background modelling framework to the stars in Table~\ref{tab:InterSamp}, using the \numax values of \citet{Lund25} for the prior setup. Only a single star failed to obtain meaningful granulation posteriors\footnote{The star $\xi$~Gem, displaying a PDS with complex activity components and generally weak oscillations in the 120~s cadence data used in this work (Appendix~\ref{app:intersamp}).}, but the remaining 53 TESS stars could be used in the present work. 

In Appendix~\ref{app:intersamp} three stars from Paper~I were found to also have interferometric radii determinations. The first, KIC~6106415, was Perky; already excluded from the pool of stars as our reference. The second, KIC~6225718, was already in the seismic sample, but we chose to replace it with the interferometric source instead. The last, KIC~8751420, had no asteroseismic radius from \citet{Sayeed25} and was therefore previously excluded from the seismic sample. Thus, with these two additions, the resulting sample consists of 56 stars with interferometric radii.

\subsection{Resulting subsamples}
To construct the calibration and validation subsamples, we first pooled all stars and grouped them by source: Legacy, KAGES, APOKASC~1, other seismic stars, and the interferometric sample. From each group, $50\%$ of the stars were selected at random to form the calibration sample, while the remaining $50\%$ were assigned to the validation sample. This procedure ensures that both subsets retain a similar balance of star source origins, with the aim of a comparable representation of the underlying methodological systematics.

As we have few evolved red-giant branch (RGB) stars in the sample, we manually excluded them from the calibration sample. For the present dataset this corresponds to the four stars with $R \geq 5\,R_\odot$, which were instead retained in the validation sample for testing. This was done to avoid disproportionate influence from stars residing in a distinct and sparsely populated region of the parameter space, restricting the calibration to the MS and SGB regime where our observational constraints are strongest. 

Figure~\ref{fig:KielOverview} presents an overview of the resulting subsamples and illustrates how both span comparable temperature and surface gravity regimes, with the exception of the four most evolved stars forced into the validation sample. The timescale decoupling identified in Paper~I and the resulting selection criteria in $\numax$ removes all K dwarfs from the original seismic sample (13 stars in total). The reference star Perky is located within the parameter space of the MS stars.

\section{A hierarchical modelling setup}\label{sec:Methods}
This section outlines the methodology used to investigate the connection between stellar radii and granulation. We adopt a Bayesian hierarchical approach \citep{GelmanBOOK}, ensuring that measurement uncertainties and parameter correlations are treated consistently throughout the analysis, thereby providing a statistically rigorous framework for the derivation of the granulation-radius relations. For this purpose we use the \textsc{CmdStan}\footnote{\url{https://mc-stan.org/docs/2_38/stan-users-guide/index.html}} software for Bayesian data analysis and modelling, which performs Hamiltonian Markov-Chain Monte-Carlo sampling \citep{Betancourt13} using the No-U-Turn algorithm \citep{Hoffman14}. The complete \texttt{Stan} code, configuration details, and convergence diagnostics are provided in Appendix~\ref{app:pystan}. 

We set up a relation of the form:
\begin{equation}\label{eq:radscal}
R_\mathrm{pred} = C\,A_\mathrm{gran}^{x}\,b^{y}\,T_\mathrm{eff}^{z}\,10^{\,w\,\mathrm{[Fe/H]}} .
\end{equation}
Here $R_\mathrm{pred}$ denotes the radius predicted from the granulation signals. We define and motivate the various input variables used as follows:
\begin{itemize}
    \item A constant scaling factor $C$.
    \item Total granulation amplitude $A_\mathrm{gran} = C_\mathrm{bol} \sqrt{\sum a_i^2}$. \\
    In Paper~I it was found that the total granulation amplitudes were the most consistent across different background models. Moreover, it combines the individual normalised amplitudes, $a_i$, of the various components into a single estimate. This estimate also takes into account the bolometric correction for the passbands, calculated as $C_\mathrm{bol,\kepler} = \left(\teff / 5934~\textup{K}\right)^{0.8}$ \citep{Michel09, Ballot11} for stars observed by \kepler and $C_\mathrm{bol,TESS}=\left(\teff/4714 \right)^{0.81}$ \citep{Lund19} for TESS.
    
    \item Characteristic frequency of the primary granulation component, $b$. \\
    The characteristic frequency, or equivalently the timescale $\tau=1/2\pi b$, of the strongest granulation signal in a PDS. We chose this characteristic frequency as it is well determined and clearly separated in frequency from potentially present stellar oscillations, thereby avoiding complications stemming from the adopted description of the oscillation envelope (see Sect.~4 and Appendix~E of Paper~I for details).
    
    \item Stellar effective temperature, \teff. \\
    The effective temperature sets the surface radiative flux ($F_\mathrm{bol}\propto T_\mathrm{eff}^4$) and enters the surface pressure scale height through the thermodynamic structure ($H_p \propto T_\mathrm{eff}/g$), thereby influencing the characteristic size and energetics of near-surface convective cells \citep{Nordlund09}. Including $\teff$ therefore provides an additional constraint on the underlying stellar structure, beyond what is directly encoded in the granulation amplitudes and timescales, which primarily probe the stochastic manifestation of convection rather than the global thermal stratification \citep{Kallinger2014}.

    \item Stellar surface metallicity, \FeH.\\
    It is expected that stellar metallicity affects the convective velocity flows and thereby the granulation signals (e.g. \citealt{Corsaro17,Yu18,Diaz22}). The catalogue of \citet{Sayeed25} compiles the observed spectroscopic \FeH values for the majority of the stars across our sample. How we treat the stars with and without an observed value is described in Sect.~\ref{subsec:metallicity}. Since \FeH is given in dex, we write its contribution to Eq.~\ref{eq:radscal} as $10^{w\,\FeH}$, which is mathematically equivalent to a standard power law in linear abundance ratio ($Z/Z_\odot$): $10^{w\,\FeH}=(Z/Z_\odot)^w$. As the inference will be performed in logarithmic space (Sect.~\ref{subsec:PerkyRelSpace}), the metallicity term thus takes an additive form in the inference, analogous to the other terms.    
\end{itemize}
\begin{figure}[]
    \resizebox{\hsize}{!}{\includegraphics[width=\linewidth]{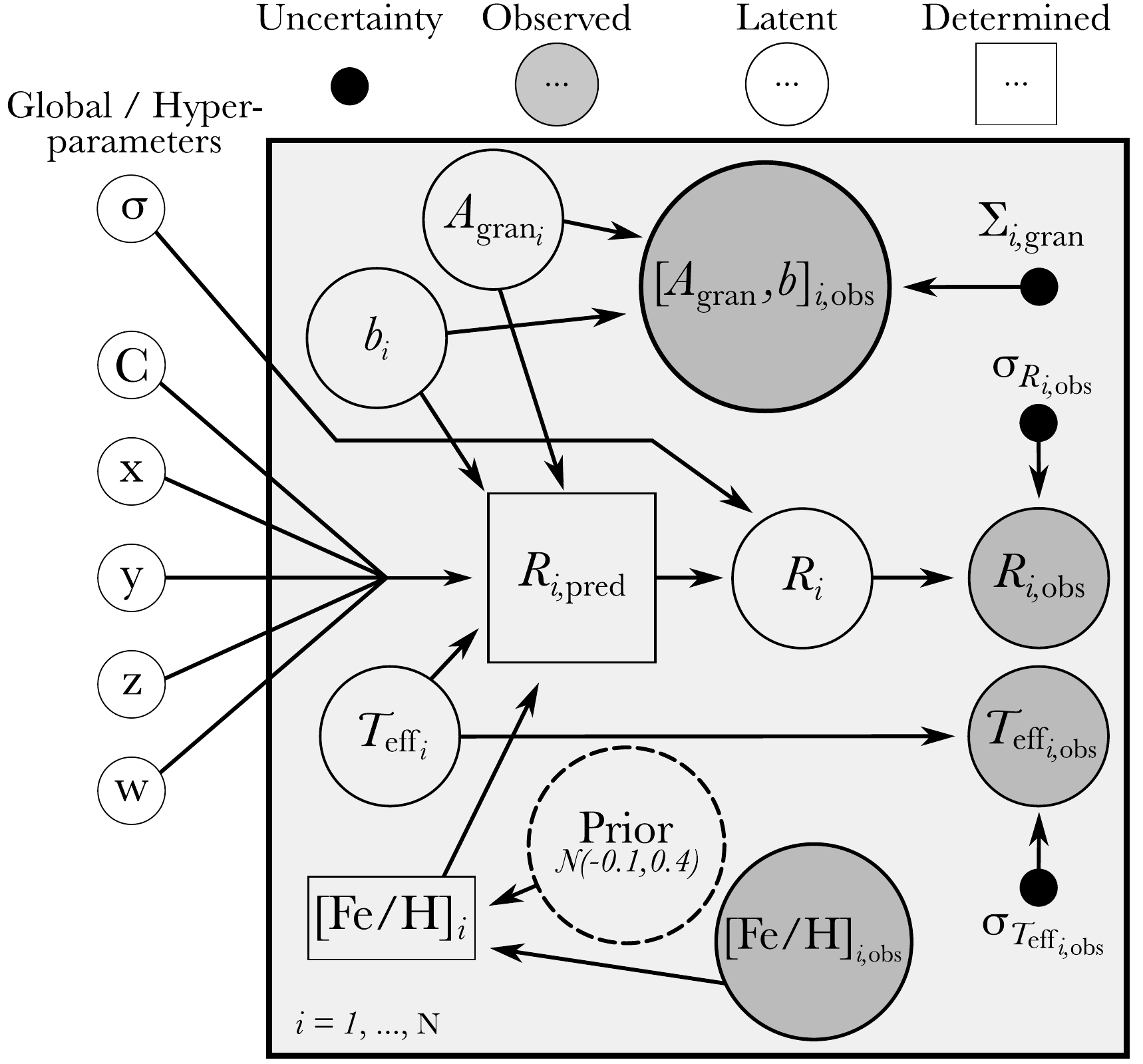}}
    \centering
    \vspace*{-2mm}
    \caption{Probabilistic graphical model for the hierarchical modelling setup. The white nodes outside the grey plane represent the global regression (hyperparameters) of the model. The circular nodes inside the grey plane are the latent variables (underlying true values) in the model, each corresponding to a shaded observation node. The observations are conditioned on their uncertainties, shown by the smaller solid nodes. If no metallicity observation exists, we draw an estimate from a prior as indicated by the dashed circle.}
    \label{fig:PGM}
    \vspace*{-2mm}
\end{figure}
Each parameter above has an associated exponent as defined by Eq.~\ref{eq:radscal}, treated as global free parameters to be determined in the inference. Importantly, three different granulation background models have been applied to all stars studied, each describing the background with a different prescription. In the following, we therefore derive a separate granulation–radius relation of the form in Eq.\ref{eq:radscal} for each model. During application in Sect.\ref{sec:Application}, this yields three radius posteriors for a given star. These are then combined in a weighted manner using the Bayesian evidences from Paper~I through Bayesian model averaging \citep{Fragoso15} to obtain a combined granulation-inferred radius posterior.

\subsection{Overview of the hierarchical probabilistic structure}\label{subsec:HierMod}
Figure~\ref{fig:PGM} shows a probabilistic graphical model for the Bayesian hierarchical modelling structure of this work. The diagram represents a directed graph of conditional probabilities, illustrating the generative statistical dependencies of the model. Accordingly, the arrows follow the generative direction. The latent parameters correspond to the underlying physical stellar properties, which give rise to the observed measurements through Gaussian likelihoods defined by the inferred uncertainties and covariances. In this representation, the observations depend on the true parameters, even though the process of inference proceeds in the opposite direction.

For each star, $i$, the hierarchical structure shown in Fig.~\ref{fig:PGM} can be interpreted as follows:
\begin{itemize}
    \item \textbf{Latent stellar properties:} 
    The nodes in Fig.~\ref{fig:PGM} represent the latent variables, that is, the underlying true stellar quantities: the total granulation amplitude ($A_\mathrm{gran})$, characteristic frequency ($b$), effective temperature ($T_\mathrm{eff}$), and radius ($R_i$).

    \item \textbf{Metallicity:} 
    Metallicity enters conditionally as an effective regression input, as described in Sect.~\ref{subsec:metallicity}, and is shown as a determined quantity in Fig.~\ref{fig:PGM}, thus entering either directly as the observed value or drawn from the indicated population prior.

    \item \textbf{Population-level relation:} 
    The parameters located outside the main plate in Fig.~\ref{fig:PGM} define a global regression that predicts a stellar radius, $R_{i,\mathrm{pred}}$, from the latent granulation and atmospheric parameters according to Eq.~\ref{eq:radscal}.

    \item \textbf{Intrinsic scatter:} 
    The true stellar radius ($R_i$) is assumed to scatter around the predicted value with an intrinsic dispersion, $\sigma$. The intrinsic dispersion is also treated as a global parameter in the model and represents the fractional $1\sigma$ uncertainty of the scaling relation due to astrophysical variability.

    \item \textbf{Observational models:} 
    The observed quantities (dark grey circles in the main plate in Fig.~\ref{fig:PGM}) are modelled as noisy measurements of their corresponding latent values. The total granulation amplitude and characteristic frequency are described by a joint likelihood that accounts for their measurement covariance, represented by $\Sigma_{i,\mathrm{gran}}$ in Fig.~\ref{fig:PGM}. This is informed by their complete marginal posterior distributions derived in Paper~I.
\end{itemize}
In this formulation, the model separates the population-level scaling relation, intrinsic astrophysical scatter, and observational uncertainties within a single, coherent hierarchical framework.

\subsection{Sampling in relative and logarithmic space}\label{subsec:PerkyRelSpace}
To improve numerical stability and sampling efficiency, the inference is not performed in absolute units. Instead, all linear observables are expressed relative to the reference star Perky, such that for $x\in\{R,\teff,A_\mathrm{gran},b\}$ we define $x_\mathrm{rel}=x/x_\mathrm{Perky}$. Because metallicity is already expressed logarithmically, it is centred by subtracting Perky’s metallicity.

The inference is then carried out in logarithmic space for the linear quantities, $\log_{10}(x_\mathrm{rel})$, with uncertainties propagated accordingly. This places the global regression quantities of the inferred granulation–radius relations on comparable scales (typically near zero) and increases computational efficiency by not having to sample across many orders of magnitude (e.g. radii in cm). Logarithmic space is also natural for strictly positive quantities such as stellar radii and for relations expected to follow a power-law, which become linear with additive scatter on a logarithmic scale (e.g. \citealt{Sivia06}). After sampling, the inferred radii are transformed back to linear units and reported in solar radii.

\subsection{Transformations of the posteriors}\label{subsec:PosTrans}
It is important that we include the correlations between the total granulation amplitudes and characteristic frequencies during our inference for the stellar radii. This is accounted for by using the marginal posterior distributions for the parameters. 

We therefore pass all marginal posterior samples ($\sim10^4-10^5$ per star) through the full chain of model-dependent transformations from fitted background parameters to the observables $A_\mathrm{gran}$ and $b$. This includes a normalisation to satisfy Parseval's theorem (see Sect.~2.2 of \citealt{Larsen2025b} for details), a rescaling depending on the sampled \numax value (see Sect.~3.1 of Paper~I), the bolometric correction using \teff \citep{Michel09, Ballot11,Lund19}, the calculation of the total granulation amplitude, and lastly the relative and logarithmic transformations. From these transformed posterior samples we compute for each star the $2\times2$ covariance matrix, $\Sigma_{i,\mathrm{gran}}$. How this matrix is used will be outlined in Sect.~\ref{subsec:MeasureModPriors}.

\subsection{The metallicity}\label{subsec:metallicity}
The stellar metallicity is part of the assumed relation in Eq.~\ref{eq:radscal}. While the catalogue by \citet{Sayeed25} gives observed metallicities for most stars, formal uncertainties are not provided, and a subset of stars have no measurements. Hence, as seen in Fig.~\ref{fig:PGM}, within the hierarchical modelling we use a determined quantity for the metallicity with no latent layer to directly inform the predicted radius.

When an observed value exists, it is used directly in the regression; when it is missing, we infer a representative metallicity from a weakly informative prior, $\mathrm{[Fe/H]} \sim \mathcal{N}(-0.1,0.4)\ \text{truncated to}\ [-4,1]\ \mathrm{dex}$. This prior peaks slightly below solar metallicity and is intentionally broad; its central 95~\% prior mass lies approximately in the range $[-0.9,0.7]$ dex. The range in available observed metallicities for our sample is roughly [-0.8, 0.5] dex. 

Thus, by allowing metallicity to be a determined quantity informed from either an observation or a prior, we simplify the modelling setup to only require a single relation for the inference, which always includes the metallicity. As we will see in Sect.~\ref{sec:Derivation}, the results we obtain are not strongly dependent on the metallicity and the choices made here.

\subsection{Measurement models and priors}\label{subsec:MeasureModPriors}
For each star, the hierarchical model introduces latent true values and links them to the measured quantities through explicit measurement models. In the description of these, we have added the subscript 'true' to the latent values seen in Fig.~\ref{fig:PGM} for the purpose of clarity. 

\begin{figure*}[t]
    \includegraphics[width=\linewidth]{Figures/DerivH.jpg}
    \centering
    \vspace*{-5.5mm}
    \caption{Results of the granulation-radius relation derivation for the two-component background model (H). \textit{Left panel:} A corner plot of the inferred global regression parameters of Eq.~\ref{eq:radscal}. The summary statistics are given as the median and 16/84 percentiles of the posteriors. \textit{Right panel:} The granulation-inferred stellar radius plotted against the reference radius for each star, using circular and triangular points for seismic and interferometric stars, respectively. In the main panel we use the inferred \teff for the prediction, while the insert shows the results obtained when using the observed \teff. The obtained fractional root-mean-square scatter, $\Delta R/R$, are shown for each. The residuals are shown below, coloured according to the granulation background model preference as indicated by the legend.}
    \label{fig:DerivH}
    \vspace*{-2mm}
\end{figure*}

For radius and effective temperature, the observed values are modelled as Gaussian draws around the latent truths, thereby constraining the underlying true values by their observations,
\begin{align*}
\log R_{i,\mathrm{obs}} &\sim \mathcal{N}\!\left(\log R_{i,\mathrm{true}},\,\sigma_{\log R,i,obs}\right), \\
\log T_{\mathrm{eff},i,\mathrm{obs}} &\sim \mathcal{N}\!\left(\log T_{\mathrm{eff},i,\mathrm{true}},\,\sigma_{\log T_\mathrm{eff},i,\mathrm{obs}}\right).
\end{align*}
The granulation observables are constrained jointly, not independently. For each star, we define the observed granulation vector as
\[
\mathbf{y}_{i,\mathrm{obs}}=
\begin{bmatrix}
\log A_{\mathrm{gran},i,\mathrm{obs}}\\
\log b_{i,\mathrm{obs}}
\end{bmatrix}.
\]
This formulation emphasizes that \(A_{\mathrm{gran}}\) and \(b\) enter as a coupled noisy measurement pair of the underlying true stellar values. Hence, the data provide an error ellipse in \((\log A_{\mathrm{gran}},\log b)\) space rather than two separate 1D constraints. The corresponding latent vector is
\[
\mathbf{y}_{i,\mathrm{true}}=
\begin{bmatrix}
\log A_{\mathrm{gran},i,\mathrm{true}}\\
\log b_{i,\mathrm{true}}
\end{bmatrix}.
\]
We then model the measurement process with a bivariate normal likelihood, implemented via Cholesky decomposition \citep{Cholesky1924},
\[
\mathbf{y}_{i,\mathrm{obs}} \sim \mathcal{N}\left(\mathbf{y}_{i,\mathrm{true}},\,\mathbf{\Sigma}_{i,\mathrm{gran}}\right),
\]
where \(\mathbf{\Sigma}_{i,\mathrm{gran}}\) is the full star-specific covariance matrix derived from the transformed marginal posteriors (Sect.~\ref{subsec:PosTrans}). In this way, correlations between \(\log A_{\mathrm{gran}}\) and \(\log b\) are preserved explicitly during inference, ensuring that only jointly plausible combinations of the two granulation observables are used. This preserves the posterior geometry from Paper~I directly in the derivation of the granulation-radius relations. Moreover, it prevents information loss by alternatively assuming them independent and reducing the marginal posteriors of the granulation observables to point-estimate summary statistics. 

The regression coefficients of the granulation-radius relations use the following priors:
\[
\log\mathrm{C}\sim\mathcal{N}(0,2),\qquad
x,y,z,w\sim\mathcal{N}(0,0.5),\qquad
\sigma\sim\mathcal{N}^{+}(0,0.2).
\]
Latent stellar quantities are also weakly informed by the measurements through broad priors centered on the observed values on a per-star basis:
\[
\log R_{i, \mathrm{true}},\ \log T_{\mathrm{eff},i,\mathrm{true}},\ \log A_{\mathrm{gran}, i, \mathrm{true}},\ \log b_{i, \mathrm{true}}
\sim
\mathcal{N}(\text{obs},\,0.5).
\]
Metallicity is treated as a direct input to the regression of Eq.~\ref{eq:radscal} by using the observed value when available and, for stars without a measurement, sampling a latent $\FeH {\sim} \mathcal{N}(-0.1,0.4)$ truncated to [-4,1] dex (Sect.~\ref{subsec:metallicity}).

\section{Deriving the granulation-radius relations}\label{sec:Derivation}
In this section we derive the granulation-radius relations using the hierarchical methodology of Sect.~\ref{sec:Methods} and the 363 stars in the calibration sample. This is done thrice-over, once for each background model H, J, and T considered in Paper~I. Hence we obtain three similar but distinct relations, but for the sake of brevity, we only show the posteriors and prediction results for the two-component model H. The corresponding results for models J and T may be found in Appendix~\ref{app:derivres}.

The left panel of Fig.~\ref{fig:DerivH} shows the corner plot of the inferred global regression parameters for the granulation-radius relation in Eq.~\ref{eq:radscal}. As we performed the inference on the parameters relative to those of Perky, the offset $C$ represents the zero-point of the relation evaluated at Perky's position. Although Perky is excluded from the calibration sample, $C\approx0$ is expected because Perky was selected to be a well-characterised, representative solar-like oscillator of the stars studied in this work -- the small inferred values confirm this choice is consistent. The parameters $x$ and $y$ represent the exponents on the total granulation amplitude and characteristic frequency, respectively. Interestingly, the constraining power of the granulation for the radius inference is carried primarily by the timescale and less so the amplitude. Yet, the sampler has indeed explored alternative possibilities, as we see a strong correlation between the two exponents for the granulation parameters. The exponent $z$, on \teff is the largest of all and thus strongly informs the inference. For the metallicity, $w$ is close to zero, thus only weakly constraining the inference. The intrinsic scatter for the derived granulation-radius relation is found to be modest near ${\approx}$4\%, reflecting a calibration sample with stars carrying different systematics by construction. 

In the right panel of Fig.~\ref{fig:DerivH} we plot the radius inferred by granulation against the reference radius for the calibration sample. All samples from the granulation and global regression parameter posteriors are passed through Eq.~\ref{eq:radscal} to form a radius posterior for each star, from which we report the median and 16th and 84th percentiles as the granulation-inferred radius. The main panel uses the latent effective temperature --- the per-star \teff inferred jointly with the other parameters of the hierarchical model, allowing the regression to account for observational noise in \teff through the latent layer. The inset repeats the comparison using the observed \teff directly, without marginalising over its underlying true value via a latent layer as in the hierarchical model, previewing the application to the validation sample in Sect.~\ref{sec:Application}. In both cases the granulation-inferred radii scatter around the 1:1 line with a root-mean-square fractional scatter of $\Delta R/R \approx 9$--$10\%$.

\begin{table}
\renewcommand{\arraystretch}{2}
\centering
\caption{Global regression parameters of Eq.~\ref{eq:radscal} inferred by the hierarchical modelling, specified for the three background models in Table~1 of Paper~I.}
\vspace*{-1mm}
\begin{tabular}{lccc}
\hline
\multicolumn{1}{c}{\thead{Parameter}} 
& \multicolumn{1}{c}{\thead{H}} 
& \multicolumn{1}{c}{\thead{J}} 
& \multicolumn{1}{c}{\thead{T}} \\
\hline
$C$ 
& $-0.018^{+0.006}_{-0.006}$ 
& $0.010^{+0.005}_{-0.005}$ 
& $-0.009^{+0.006}_{-0.006}$ \\

$x$ 
& $0.10^{+0.04}_{-0.04}$ 
& $0.11^{+0.04}_{-0.04}$ 
& $0.09^{+0.04}_{-0.04}$ \\

$y$ 
& $-0.63^{+0.02}_{-0.02}$ 
& $-0.57^{+0.02}_{-0.02}$ 
& $-0.63^{+0.02}_{-0.02}$ \\

$z$ 
& $1.20^{+0.09}_{-0.09}$ 
& $0.97^{+0.08}_{-0.08}$ 
& $1.19^{+0.08}_{-0.09}$ \\

$w$ 
& $0.05^{+0.01}_{-0.01}$ 
& $0.04^{+0.01}_{-0.01}$ 
& $0.05^{+0.01}_{-0.01}$ \\

$\sigma$ 
& $0.039^{+0.002}_{-0.002}$ 
& $0.039^{+0.002}_{-0.002}$ 
& $0.037^{+0.002}_{-0.002}$ \\
\hline
\end{tabular}
\tablefoot{The parameter estimates are given as the median and 16th and 84th percentiles of the posteriors produced by the hierarchical modelling, as seen in the corner plots of Figs.~\ref{fig:DerivH}, \ref{fig:DerivJ}, and \ref{fig:DerivT}, respectively for models H, J and T.}
\label{tab:GlobParams}
\vspace*{-2.5mm}
\end{table}

Table~\ref{tab:GlobParams} summarises the regression parameters for the granulation-radius relations for models H, J, and T. The regression parameters overlap within the uncertainties in most cases, particularly when comparing the two individual-component models H and T. The variation is larger when the nature of the background model changes to the hybrid model J, which has a single granulation amplitude but two characteristic frequencies. Here the exponent of the granulation timescale is slightly reduced and, notably, the exponent on temperature is lower by a difference of $\approx2\sigma$. However, upon inspecting the predictions in Figs.~\ref{fig:DerivJ} and \ref{fig:DerivT}, the obtained root-mean-square scatter is unchanged across the three background models. 

We assessed the sensitivity of the framework to the random sample creation in Sect.~\ref{sec:CalibValid} by rerunning the full analysis for 10 different random seeds, each producing a unique division of the stars between the calibration and validation samples. The random changes cause slight variations in the present systematics within the calibration sample that we derive the methodology upon. Overall, the results indicate that the framework is robust to the choice of random seed, with only small across-seed scatter in the global regression parameters and in the standardised residuals (quantified in Sect.~\ref{subsec:AppValid}).

\begin{figure*}[t]
    \includegraphics[width=17cm]{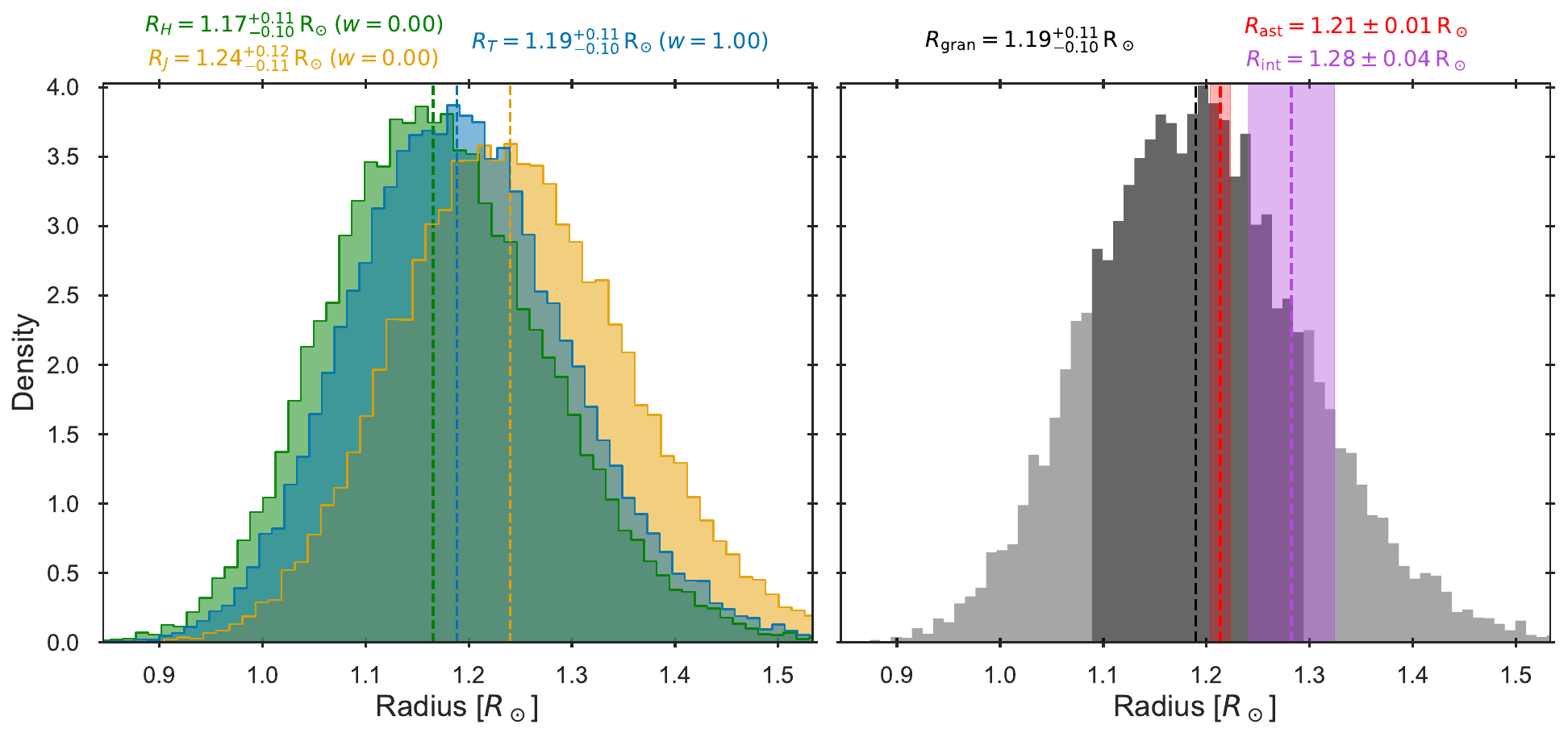}
    \centering
    \vspace*{-1mm}
    \caption{Separate granulation background model posteriors (left) and the combined Bayesian model averaged radius posterior (right) for the reference star Perky. The vertical dashed lines show the median values of the radius posteriors, with the colours corresponding to the legends. The red and purple vertical dashed lines in the right hand plot shows the asteroseismic and interferometric radius of Perky, respectively, with the uncertainties indicated by the corresponding shaded bands. The 16 and 84 percentiles of the combined radius posterior in the right panel are indicated by the darker shaded region.}
    \label{fig:PerkyPos}
    \vspace*{-2mm}
\end{figure*}

\section{Applying the granulation-radius relations}\label{sec:Application}
In the following we will outline the workflow for applying the derived granulation-radius relations to a single star, before using and evaluating it across the validation sample. We infer the radius independently with each granulation background model and then combine those inferences with Bayesian model averaging \citep{Fragoso15}. Let $\mathcal{M} \in \{H,J,T\}$ denote the model index (two-, hybrid-, and three-component granulation formulations, respectively). For a given star, $i$, the set of observables is 
\begin{equation}
    \mathbf{d}_{i,\mathcal{M}}=\left(A_{\mathrm{gran},i,\mathcal{M}},\, b_{i,\mathcal{M}},\,T_{\mathrm{eff},i},\,\mathrm{[Fe/H]}_i\right).
\end{equation}
Here, $A_{\mathrm{gran},i,\mathcal{M}}$ and $b_{i,\mathcal{M}}$ are the granulation observables of the specific background model $\mathcal{M}$, obtained from the transformed samples as described in Sect.\ref{subsec:PosTrans}. The atmospheric parameters \teff and \FeH are shared across models $\mathcal{M}$ and enter the radius inference as Gaussian draws about their measured values. Similarly to the derivation stage in Sect.\ref{sec:Derivation}, we treat metallicity as a determined quantity. For stars with a measured metallicity, we draw \FeH from a Gaussian centred on the observed value, assuming a conservative uncertainty of $0.2,\mathrm{dex}$, as formal uncertainties are not provided in \citealt{Sayeed25}. For stars without a measurement, we instead draw from the population prior $\mathrm{[Fe/H]}_i \sim \mathcal{N}(-0.1,0.4)$. This ensures that metallicity information, or its absence, is consistently propagated into the inference.

Using the samples from the marginal posteriors of the regression coefficients in Table~\ref{tab:GlobParams} for model $\mathcal{M}$ and the observable set $\mathbf{d}_{i,\mathcal{M}}$, we evaluate the granulation-inferred radius of each posterior draw when passed through Eq.~\ref{eq:radscal}. Each posterior draw combines: (i) one draw of the derived regression coefficients, (ii) one draw of the star’s observables, and (iii) one draw of the intrinsic granulation-radius relation scatter $\sigma$. This yields a model-specific set of posterior samples
\begin{equation}\label{eq:radpos_permod}
    \left\{R_{i,\mathcal{M}}^{(s)}\right\}_{s=1}^{N_s} \sim p(R_i \mid \mathbf{d}_{i,\mathcal{M}}),
\end{equation}
with $s$ indexing the 26 664 posterior draws (see Appendix~\ref{app:pystan}).

A key point is that the granulation observables are also not sampled independently during the application stage. As in the derivation setup (Sect.~\ref{subsec:MeasureModPriors}), for each star and model $\mathcal{M}$, we propagate their joint posterior covariance by sampling from the corresponding marginal posteriors. This preserves the observed correlation between $A_{\mathrm{gran},i,\mathcal{M}}$ and $b_{i,\mathcal{M}}$ in the radius inference. We also include the intrinsic scatter term of the derived relations, \(\sigma_\mathcal{M}\), by adding a random draw consistent with that model’s inferred scatter at each posterior realisation. Therefore, the width of $p(R_i\mid \mathbf{d}_{i,\mathcal{M}})$ reflects both measurement uncertainties through the use of the granulation posteriors and atmospheric parameters, and also the predictive precision of the derived granulation-radius relations themselves.

To combine models, we use the estimated Bayesian evidences obtained from the nested-sampling \citep{Skilling04} of Paper~I. Using the evidence, $Z_\mathcal{M}$, for a given model the Bayesian model-averaged weight becomes,
\begin{equation}
    w_\mathcal{M}=\frac{Z_\mathcal{M}}{\sum_{k\in\{J,H,T\}}Z_k},
\end{equation}
where $k$ is a summation index over all background models. The weights satisfy $\sum w_\mathcal{M}=1$, and they can therefore be interpreted as relative model support for a given star.

The final radius posterior is then the mixture
\begin{equation}\label{eq:radpos}
    p(R_i\mid \mathbf{d}_i)=\sum_{\mathcal{M}\in\{J,H,T\}} w_\mathcal{M}\,p(R_i\mid \mathbf{d}_{i,\mathcal{M}}).
\end{equation}
In Eq.~\ref{eq:radpos} we have implicitly assumed the prior-odds ratio between the three models to be unity, that is, we assume no prior preference between the background models (see Appendix~A of \citealt{Larsen2025b} for a thorough description). In practice, we generate this distribution by weighted resampling from the three model-specific posterior sample sets of Eq.~\ref{eq:radpos_permod}. We report the median of the combined posterior in Eq.~\ref{eq:radpos} as the granulation-inferred radius,
\begin{equation*}
    R_{\mathrm{gran}}=\mathrm{median}\!\left[p(R_i\mid \mathbf{d}_i)\right],
\end{equation*}
and use the 16th and 84th percentiles as the credibility interval. In summary, the procedure for a single star is as follows:
\begin{enumerate}
    \item Generate a model-specific radius posterior $p(R_i\mid \mathbf{d}_{i,\mathcal{M}})$ for each $\mathcal{M}\in\{J,H,T\}$ using the respective derived granulation-radius relation.
    \item Compute evidence-based weights $w_\mathcal{M}$ from $\log Z_\mathcal{M}$.
    \item Form the Bayesian model-averaged posterior (Eq.~\ref{eq:radpos}) as the weighted mixture across background models.
    \item Report $R_{\mathrm{gran}}$ (median) and $(R_{16},R_{84})$ as credibility interval.
\end{enumerate}

\subsection{Application to Perky}
The first star we applied the above methodology to was the Legacy star Perky. As outlined in Sect.~\ref{subsec:PerkyRelSpace}, it is the reference star used within this framework. As such, it has not been used to derive the granulation-radius relations in Sect.~\ref{sec:Derivation}, yet remains a well-characterised seismic star for us to validate the application on.

Figure~\ref{fig:PerkyPos} shows the obtained model-specific radius posteriors from Eq.~\ref{eq:radpos_permod} and the resulting combined posterior of Eq.~\ref{eq:radpos}. The granulation background models produce radius posteriors which overlap, all approximately Gaussian in shape. Combining them through Bayesian model averaging -- where for Perky model T dominates entirely in the model preference -- we obtain a radius posterior with a median within 2\%  of the asteroseismic radius. Comparing to the interferometric radius of Perky, the granulation differs by almost 8\%. However, the 16th and 84th percentiles encompass both reference radii for Perky -- an overlap not present between the asteroseismic and interferometric radius estimates. 

The granulation-inferred radius for Perky reaches a precision of approximately 9\%, compared to the ${\approx}1$--$2\%$ achievable with asteroseismology and the ${\approx}3\%$ of interferometry, reflecting the broader combined posterior in this method. Nevertheless, when applied to a well-characterised star with high-quality observations, the method recovers the reference radii to within the uncertainties. This level of agreement provides an indication that the methodology captures the relevant physical information for constraining the stellar radius. This is consistent with its design to remain robust when trained and applied across heterogeneous samples spanning a wide range of systematics.

\begin{figure*}[t]
    \includegraphics[width=17cm]{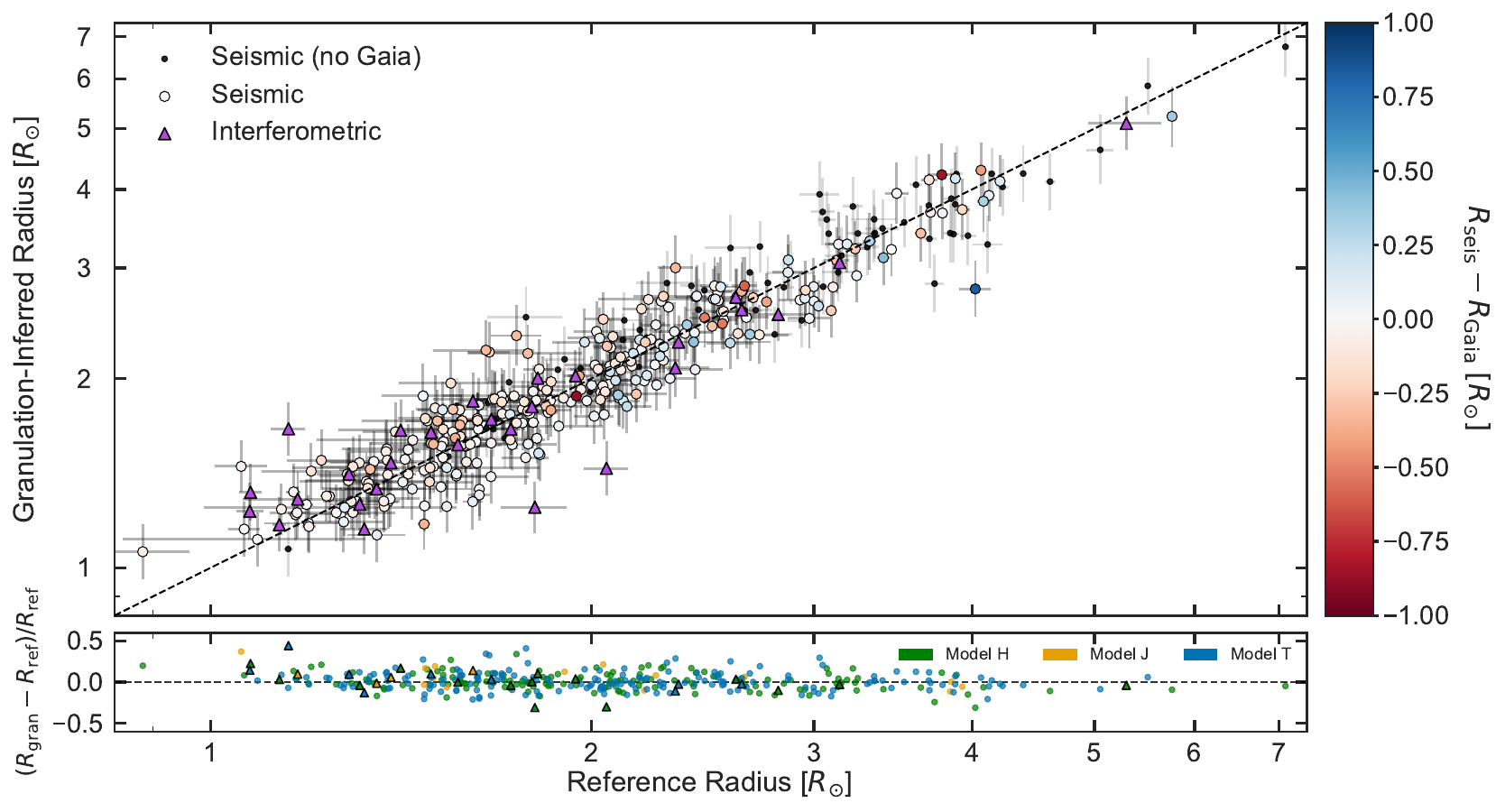}
    \centering
    \vspace*{-3mm}
    \caption{Granulation-inferred radii compared to the reference radii for the validation sample. The stars with asteroseismic and interferometric radii are plotted with circular and purple triangular points, respectively. Where available, the seismic stars are colour-coded by the difference between the seismic and Gaia radius, and left black otherwise. Note that they are plotted in ascending order, such that those with the largest differences are seen on top, while the majority which are in close agreement lie underneath. The residuals are shown below, coloured according to granulation background model preference as indicated by the legend.}
    \label{fig:ApplicationResults}
    \vspace*{-2mm}
\end{figure*}

\subsection{Application to the validation sample}\label{subsec:AppValid}
In this section, we assess the statistical performance of the granulation-inferred radii by applying the developed framework to the 367 stars in the validation sample. The diagnostics used for this purpose are built around the distribution of standardised residuals (z-scores), which jointly tests the accuracy and calibration of the predictions and their associated uncertainties. Two complementary quantities are drawn from this distribution: (i) the empirical 1-$\sigma$ coverage -- i.e. the fraction of stars with $|z_i| < 1$ -- and (ii) the shape of the complete z-score distribution, which reveals whether the uncertainties are appropriately scaled and whether any systematic bias is present.

Under correct calibration and Gaussian errors, the z-scores should follow an $\mathcal{N}(0,1)$ distribution (e.g. \citealt{Wasserman04,Sivia06}). For each star, we define the standardised residual as 
\begin{equation}
    z_i=\frac{R_{\mathrm{gran},i}-R_{\mathrm{ref},i}}{\sqrt{\sigma_{\mathrm{gran},i}^2+\sigma_{\mathrm{ref},i}^2}},
\end{equation}
where $\sigma_{\mathrm{gran},i}$ is obtained by symmetrising the posterior credibility interval and $\sigma_{\mathrm{ref},i}$ denotes the reference radius uncertainty. The quantity $z_i$ therefore measures the discrepancy between the two radius estimates in units of the total expected uncertainty. By construction, $z_i = 0$ corresponds to perfect agreement, $|z_i| = 1$ indicates a difference at the one-sigma level, and larger values of $|z_i|$ reflect increasingly significant tension.

\begin{figure}[]
    \includegraphics[scale=0.63]{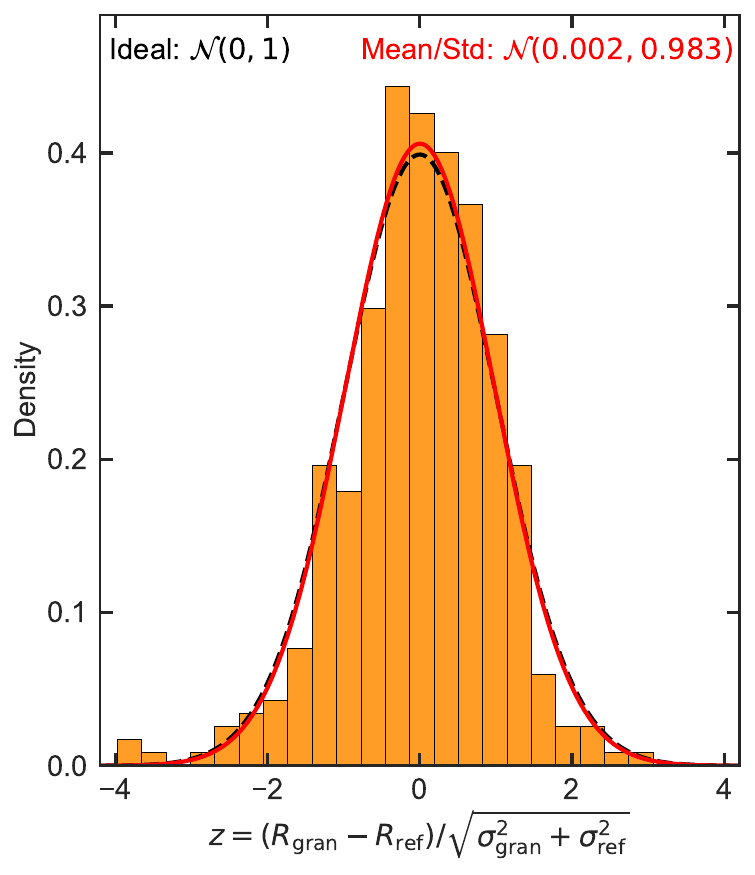}
    \centering
    \vspace*{-2.5mm}
    \caption{Distribution of the standardised residuals (z-scores). The ideal distribution is shown by the black dashed profile, while a Gaussian fit to the z-scores reporting the mean and standard deviation metrics is given by the red profile.}
    \label{fig:Zscores}
    \vspace*{-3mm}
\end{figure}

The granulation-inferred radii of the validation sample are shown in Fig.~\ref{fig:ApplicationResults}. We recover the reference radii within 1-$\sigma$ for ${\approx}73\%$ of the stars, indicating that the granulation-inferred radius estimates are statistically consistent with the reference values for the majority of the sample. This fraction exceeds the nominal ${\approx}68\%$ expected under idealised Gaussian coverage, implying that the posterior intervals inferred from Eq.~\ref{eq:radpos} are, if anything, slightly conservative. This results in a mean radius precision of ${\approx}10\%$ across the validation sample. Within stellar grid-based modelling, radius estimates reach a precision of ${\approx}20\%$ when based on spectroscopic constraints alone, improving to ${\approx}5$--$8\%$ when global asteroseismic parameters are included (see e.g. Fig.~8 of \citealt{Aguirre22}). The granulation–radius method therefore falls between these regimes.

The z-score distribution is shown in Fig.~\ref{fig:Zscores} and supports the above interpretation. A Gaussian fit yields a dispersion close to unity, demonstrating that the uncertainties are appropriately scaled overall. The distribution exhibits a mean close to zero, which means the methodology delivers radii with no systematic offsets from the reference radii, highlighting the importance of a heterogeneous calibration sample when the aim is general applicability.

To assess the robustness of this result, similarly to the derivation step in Sect.~\ref{sec:Derivation}, we inspected how the centroid and dispersion of the z-score distribution changed across 10 random sample creation seeds. The centroid consistently lay near 0 with a mean variation of 0.02, and the dispersion near 1 with a mean variation of 0.06, showing that the developed methodology is robust against the random selection of calibration and validation samples.

\subsection{Comparing to Gaia radii}\label{subsec:GaiaComp}
In Fig.~\ref{fig:ApplicationResults} we colour-coded the 216 seismic stars from the validation sample with available Gaia radii \citep{GaiaDR3_AstParamRelease} by the difference in the radius estimates. No clear trend is seen in the distribution of these differences about the 1:1 line, although individual discrepancies reach up to $\pm 1R_\odot$. Gaia delivers radius estimates for vast numbers of stars in its current and forthcoming data releases \citep{GaiaDR3,GaiaDR3_AstParamRelease}. We therefore assess how the precision and systematic behaviour of Gaia radii compare to our granulation-inferred estimates. 

We repeated the application of our methodology for the 216 stars in the validation sample using the Gaia radii as the reference. A clear difference emerges in the standardised residuals across all 10 random-seed samples: the distributions exhibit a systematic offset in their centroid, with a mean of $-0.17$. This implies that the Gaia radii are systematically lower than our granulation-inferred radii by approximately $0.043$ $\mathrm{R}_\odot$ in absolute terms. Given that the reported precision of Gaia radii in the temperature and radius regime of the validation sample is of order $0.025$ $\mathrm{R}_\odot$, this offset is significant. The granulation-radius relations thus provide an independent and complementary route to estimating stellar radii. In this role, they can also serve as a useful consistency check on the Gaia-derived radii.

\section{Current status and future prospects}\label{sec:Discuss}
In this work we have established a method for inferring stellar radii directly from granulation signals in long-duration photometric time series. The availability of reliable stellar radii for field stars is widely applicable, extending into exoplanetary studies of planet composition and demographics. Because surface granulation is a generic property of stars with outer convective envelopes, the underlying observable on which this method rests is widely present across the Hertzsprung–Russell diagram. In that sense, the approach is intrinsically general. The applications throughout Sect.~5 presented the predictive performance of the granulation-radius relations. The methodology was robust to changes in calibration and validation samples, and showed accurate recovery of the reference radii. A further avenue to improve robustness for general application of the method would be to incorporate independent luminosity constraints from Gaia parallaxes and photometry \citep{GaiaDR3} as additional information within the hierarchical inference.

A limitation, however, follows from the decoupling between granulation and oscillation timescales identified in Paper~I for cool MS stars (K- and M-dwarfs), showing that in this regime the granulation timescales change behaviour. This makes it difficult to use them as a diagnostic for the global stellar properties in a predictable manner, thus rendering the present formulation unsuitable for the lower MS. The derived granulation–radius relations therefore provide an independent route to stellar radii primarily in the domain where asteroseismology is, in principle, also available. The essential distinction is that oscillation modes need not be individually detectable for granulation to be measurable. This widens the practical applicability to stars whose power spectra contain a measurable background signal but lack oscillations that are sufficiently visible to allow precise seismic radius estimation. In particular, magnetic activity and rotation are known to suppress and broaden the oscillation power excess, complicating robust asteroseismic inference, while not necessarily precluding the measurement of the granulation background. Such applications would extend radius estimates to a broader subset of photometric survey targets and offers a complementary diagnostic to asteroseismic approaches.

Moreover, the granulation-inferred radii are comparatively model-independent as they are obtained directly from observed granulation properties, drawing on the surface signals of convection rather than the oscillations. In contrast, the inferred parameters obtained by forward or grid-based asteroseismic modelling using stellar evolution models necessarily depend on the adopted input physics and modelling assumptions (e.g. \citealt{Kippenhahn13,Aguirre20,Joyce23,Ong25}). In cases where the oscillation signal is challenging to interpret, stellar properties are often inferred directly from global seismic observables (i.e. $\nu_{\max}$ and $\Delta\nu$) via the asteroseismic scaling relations. These are known to exhibit biases for stars that differ from the Sun in metallicity or evolutionary state, reflecting the underlying assumptions of the relations (e.g. \citealt{Viani17, Larsen25, Lundkvist25, Hekker26}). Similarly, interferometric radius estimates rely on converting measured uniform-disc angular diameters into limb-darkened values using stellar atmosphere models, and are therefore sensitive to the adopted limb-darkening prescription (e.g. \citealt{Davis2000,Neilson12}). During derivation of the granulation-radius relations we do inherit some of this model dependence, but the hierarchical framework mitigates it: reference radii enter as noisy observations of a latent true value rather than direct structural constraints, and the calibration sample was constructed to be heterogenous with diverse input physics, thereby diluting pipeline-specific biases rather than being coherently propagated into the resulting granulation-radius relations. The application to the validation sample presented in Sect.~\ref{sec:Application} supports this reasoning, with the z-score distribution showing no detectable systematic offset. 

What the method remains sensitive to, however, is the stellar parameter space covered by the calibration sample. The obtained regression parameters of Table~\ref{tab:GlobParams} are valid for stars representative of this sample, and extrapolating beyond its regime -- for example, to very metal-poor stars, evolved giants, or massive stars -- is not guaranteed to produce reliable results. In addition, if a star with an unobserved metallicity lies far from the calibration regime near solar metallicity, the \FeH population prior will bias us toward near-solar values. For TIC~202872681 with a metallicity of $\FeH = -3.96\pm0.09$ dex \citep{Placco24}, a star lying ${\sim}9\sigma$ below the prior centroid, omitting the observed \FeH shifts the granulation-inferred radius on a $1.7\sigma$ level. Centring the prior on the observed metallicity restores agreement, confirming that the modest shift is prior-driven rather than a failure of the granulation-radius relations.

More generally, it is expected that the granulation observables carry a partial surface-gravity signal, given that the characteristic frequency $b \propto \nu_\mathrm{max} \propto g/\sqrt{T_\mathrm{eff}}$. If all the constructed model did was to predict \logg and invert it under $g = GM/R^2$, this would introduce a sensitivity to the mass distribution of the calibration sample. Appendix~\ref{app:PredConstraints} shows that at most $39\%$ of the variance in the obtained residuals (Fig.~\ref{fig:logg_rad_const}) may be described by surface-gravity constraints. The constructed hierarchical model thereby retrieves actual radius-specific information from the granulation observables, mitigating the dependence on the mass distribution of the calibration sample.

Taken together, these effects indicate that the method remains influenced by how well a target star is represented within the parameter space spanned by the calibration sample, as well as by the residual sensitivity to the sample’s underlying mass distribution. We therefore advise caution if considering broad applications to stars not representative of the calibration sample.

\subsection{Assessing evolutionary dependence and prospects for red giants}
A central question concerns the extent to which the granulation–radius relations depend on stellar evolutionary phase. By construction of the catalogue in \citet{Sayeed25}, it is dominated by MS and SGB stars, with only limited representation of more evolved members.

To probe evolutionary sensitivity within the available range, we re-derived the relations after partitioning the calibration sample at $\numax \simeq 1000\,\mu$Hz into an MS-only subsample. Across all three background prescriptions, the recovered global regression parameters (Table~\ref{tab:GlobParams}) remained largely consistent showing variations close to the $1\sigma$ level, with the exception of the effective-temperature exponent $z$. For the MS-only subsample, $z$ increases at roughly the $3\sigma$ level relative to the full calibration sample. While statistically noticeable, this shift did not alter the overall predictive behaviour: the subsequently obtained granulation-inferred radii for the MS stars in the validation sample recovered the reference radii to a similar degree of accuracy, showing an overlap within $1\sigma\approx72\%$ of the time. Notably, the centroid and dispersion of the standardised residuals remained close to the ideal $\mathcal{N}(0,1)$.

Extending this investigation to stars on the red-giant branch by re-deriving the granulation-radius relations using a sample of evolved stars is a planned future effort. Large catalogues characterising thousands of red giants observed by \kepler and TESS are available (e.g. \citealt{Hon18,Hon22,Pinsonneault24}). It will require substantial computational investment, but repeating the background modelling of Paper~I and the granulation-radius relation derivations in the present work for thousands of \kepler red giants is feasible. This will aid in solidifying the method across the evolutionary phases of low-mass stars and shed further light on a tentative evolutionary dependence of the granulation-radius relations. Moreover, core helium-burning red-clump stars present a promising application, as their complex interior structure leads to oscillation spectra that are challenging to model asteroseismically \citep{Schimak26}.

\subsection{Granulation-based radii with PLATO}
The upcoming ESA PLATO mission \citep{Rauer25} will obtain high-precision photometric time series for hundreds of thousands of stars, with a primary focus on bright MS and SGB targets \citep{Montalto21}. For the subset of stars exhibiting clearly detectable solar-like oscillations, expected to exceed $70\%$ of observed MS stars with $M<1.2M_\odot$, PLATO aims to deliver asteroseismic radii with uncertainties below the $2\%$ level \citep{Goupil24}. The resulting catalogue of stars with precisely characterised radii will represent a substantial enlargement to that currently available to the community.

Such an enlarged calibration set offers a natural opportunity to revisit and refine the granulation-radius relations. The hierarchical framework constructed here -- regressing $\log R$ against total granulation amplitude, primary characteristic frequency, effective temperature, and surface metallicity with propagation of the granulation posterior in the inference -- is inherently scalable. A re-derivation on the PLATO seismic catalogue would benefit simultaneously from the sheer number of stars, the wider coverage of stellar types, and the internal homogeneity of a single space-based survey. In combination, these effects would be expected to tighten the posterior constraints on the regression coefficients $(C, x, y, z, w)$ and reduce the inferred intrinsic scatter $\sigma$. This would translate directly into improved per-star precision beyond the present $\approx$10\% level.

Equally important is the complementary population that PLATO will observe: stars for which oscillation modes are not individually resolved or confidently detected. In such cases, PLATO will either rely on global asteroseismic observables or not return a seismic radius. As discussed previously, granulation, as a stochastic surface phenomenon, may remain measurable even when robust seismic diagnostics are difficult to obtain, thereby offering a complementary estimate of the stellar radius. A recalibrated framework based on the expanded PLATO seismic sample would therefore provide an additional independent and internally consistent channel for radius estimation, extending its applicability to a broader subset of stars where seismic constraints are limited or uncertain.

The implications extend to exoplanet parameter estimation. Transit observations constrain only the planet-to-star radius ratio; the absolute planetary radius -- and thus possible inferences on composition or habitability -- depends on the stellar radius. Many of the PLATO targets, with or without a seismic radius, are expected to host transiting planets \citep{Matuszewski23}. In either case, a pipeline for granulation-inferred radii operating on the same photometric data used for planet detection may offer a potential auxiliary source of stellar information, provided the granulation signal remains sufficiently constrained. Where a seismic radius is available, a granulation-inferred radius provides an independent, photometry-based cross-check derived from the same data. Where seismic radii are unavailable or unreliable, Gaia radii will often still be accessible, and as discussed in Sect.~\ref{subsec:GaiaComp}, granulation-inferred radii may serve as an independent comparison in that regime as well. In this way, granulation-inferred radii could contribute to the scientific return of large photometric surveys by extracting additional information from an otherwise auxiliary background signal.

\section{Conclusion}\label{sec:Conclusion}
This study set out to quantify how observations of the convection-driven granulation signatures encode fundamental stellar properties, focusing on stellar radius. To this end, we constructed a Bayesian hierarchical framework to derive granulation-radius relations by building directly on the granulation posteriors from Paper~I, combining them with the atmospheric and asteroseismic characterisations compiled by \citet{Sayeed25}, and further supplementing these with a sample of TESS stars with interferometric radius estimates. The result is a self-consistent methodology for inferring stellar radii directly from granulation signatures in long-duration space-based photometry, applicable to data from \kepler, TESS, and the upcoming PLATO mission. Importantly, while reliable inference requires confident detection of the primary granulation component, the method does not require the detection of asteroseismic oscillations. It may therefore, in principle, yield radius estimates where seismic patterns are unresolved or otherwise unreliable — for example, under strong magnetic suppression or rapid rotation — transforming an otherwise background signal into a quantitative stellar diagnostic. The principal results of this work are summarised as follows:
\begin{itemize}
    \item We employed Bayesian hierarchical modelling to perform a regression in log-space (Eq.~\ref{eq:radscal}) between stellar radius and the observables: total granulation amplitude, the primary characteristic granulation frequency, effective temperature, and surface metallicity. Separate relations were derived for each of the three granulation background models introduced in Paper~I, yielding the global regression parameters listed in Table~\ref{tab:GlobParams}. The framework explicitly incorporates the intrinsic scatter, $\sigma$, of the derived granulation-radius relations, while consistently propagating both measurement uncertainties and the full morphology of the marginal posteriors for the granulation parameters. Finally, Bayesian model averaging is used to combine the results, producing a radius posterior weighted by the relative suitability of each background model.
    
    \item Applied to the independent validation sample (Sect.~\ref{sec:Application}), the granulation-inferred radii recovered the reference radii within $1\sigma$ in ${\approx}73\%$ of cases. The achieved precision of the granulation-inferred radii were of the order ${\approx}10\%$. Compared to grid-based stellar modelling, this constitutes a substantial improvement over estimates based solely on spectroscopic constraints (${\approx}20\%$), is comparable to those obtained using global asteroseismic parameters (${\approx}5$--$8\%$), and remains less precise than modelling based on individual oscillation frequencies (${\approx}1$--$2\%$). The granulation-radius relations therefore occupy a complementary regime of precision while offering wide applicability.
    
    \item The methodology is constructed and calibrated on heterogeneous data spanning different instruments, analysis pipelines, and data qualities, thereby accounting directly for different uncertainties and systematics inherent in the datasets. This ensures that the inferred granulation-radius relations remain robust when applied to diverse observational data, supporting their use as a generally applicable tool for stellar characterisation across large and varied samples.
\end{itemize}
The present formulation is not applicable to cool K- and M-dwarfs, where Paper~I demonstrated a decoupling between granulation and oscillation timescales. Rather than a mere limitation, this behaviour points to a physical transition in how surface convection reflects global stellar structure. Future work extending the calibration to red-giant stars, incorporating external constraints such as Gaia-based luminosities, and exploiting the vastly enlarged calibration samples expected from PLATO, will allow the granulation-radius relations to be re-derived with greater precision and broader evolutionary coverage. As photometric surveys continue to expand in scale and quality, granulation-inferred radii offer a scalable and internally consistent pathway to stellar characterisation -- transforming a ubiquitous surface phenomenon into a robust tool for studying stellar structure across large and diverse populations.

\begin{acknowledgements}
    JRL wishes to thank the members of SAC in Aarhus and the Sun, Stars and Exoplanets group in Birmingham for comments and discussions regarding the paper. The authors thank Mikkel N. Lund for aiding with the data acquisition for the interferometric sample in Appendix A. This work was supported by a research grant (42101) from VILLUM FONDEN. MSL acknowledges support from The Independent Research Fund Denmark's Inge Lehmann  program (grant  agreement  no.:  1131-00014B). Funding for the Stellar Astrophysics Centre was provided by The Danish National Research Foundation (grant agreement no.: DNRF106). The numerical results presented in this work were partly obtained at the Centre for Scientific Computing, Aarhus \url{https://phys.au.dk/forskning/faciliteter/cscaa/}. This paper received funding from the European Research Council (ERC) under the European Union’s Horizon 2020 research and innovation programme (CartographY GA. 804752). This paper includes data collected with the \kepler and TESS missions, obtained from the MAST data archive at the Space Telescope Science Institute (STScI). Funding for the Kepler mission was provided by the NASA Science Mission Directorate. Funding for the TESS mission is provided by the NASA Explorer Program. STScI is operated by the Association of Universities for Research in Astronomy, Inc., under NASA contract NAS 5–26555. This work presents results from the European Space Agency (ESA) space mission Gaia. Gaia data are being processed by the Gaia Data Processing and Analysis Consortium (DPAC). Funding for the DPAC is provided by national institutions, in particular the institutions participating in the Gaia MultiLateral Agreement (MLA). The Gaia mission website is \url{https://www.cosmos.esa.int/gaia}. The Gaia archive website is \url{https://archives.esac.esa.int/gaia}.
\end{acknowledgements}

\section*{Data availability}
The methodology for the derivation and application of the granulation-radius scaling relations is planned for future release on Github, in an integrated pipeline with the background modelling framework of Paper~I. For now, preliminary access to both may be granted by reasonable request to the first author.

\bibliographystyle{aa.bst} 
\bibliography{bibliography.bib} 

\newpage
\begin{appendix}
{\raggedbottom
\section{Constructing the interferometry sample}\label{app:intersamp}
For the construction of the calibration and validation samples, we required stars with measured interferometric radii and photometric time series. For this purpose we turned to the work of \citet{Lund25}, which compiles the brightest solar-like oscillators observed by the TESS \citep{Ricker14}, all visible with the naked eye ($V_\mathrm{mag}\leq 6$).

First, however, we searched the catalogue of \citet{Soubiran24} for any of our seismic stars from Paper~I. Three stars were found: KIC~6106415 (Perky), KIC~6225718, and KIC~8751420. The latter of these was studied in Paper~I, however it did not have a seismic radius, effective temperature, or surface metallicity available in \citet{Sayeed25}, and was therefore discarded from the seismic sample in Sect.~\ref{subsec:seissample}. Yet, from \citet{Soubiran24} we retrieved $\teff = 5268\pm18$~K and $\FeH = -0.11$~dex for KIC~8751420, such that it may be adopted into the interferometric sample instead. For all three stars, the retrieved angular diameters and Hipparcos parallaxes \citep{vanLeeuwen07} are listed in Table~\ref{tab:Inter3adds}.

Table~D.4 of \citet{Lund25} provides the stars in their catalogue with interferometric measurements. Using this table, we selected the 50 stars displaying $200 < \numax < 3000$~\muHz and, as they are bright, adopted the Hipparcos parallaxes. From the respective interferometric sources given in the table, we retrieved the limb-darkening corrected angular diameters from the most recent works. To the extent possible, we recorded the \teff and \FeH estimates from the same source, turning to alternatives when not provided. Next, we searched for additional stars satisfying the same selection criteria in the catalogue of \citet{Soubiran24}, accepting those with multi-sector TESS data available, which resulted in 4 additional stars. To construct the light curves of these 4 additions, the TESS target pixel files were retrieved and custom aperture photometry performed following \citet{Lund15} and \citet{Lund25}. The PDSs were computed following \citet{Handberg11}. For all the 54 TESS stars we used 120~s cadence data. Table~\ref{tab:InterSamp} compiles the data retrieved and their respective sources. 

We thus have 57 stars with available interferometric measurements. The interferometric radii of all targets in Tables~\ref{tab:Inter3adds} and \ref{tab:InterSamp} were calculated from the limb-darkened angular diameter $\theta_\mathrm{LD}$ and the Hipparcos parallax $\varpi$ as,
\begin{equation}
    \frac{R_\mathrm{int}}{R_\odot} = \frac{\theta_\mathrm{LD}}{2\,\varpi} \cdot \frac{1\,\mathrm{AU}}{R_\odot} \, ,
    \label{eq:Rint}
\end{equation}
where $\theta_\mathrm{LD}$ and $\varpi$ are both in milliarcseconds, so their ratio yields the physical radius in astronomical units (AU), which is then converted to solar radii. The uncertainty on $R_\mathrm{int}$ was obtained by propagating the independent measurement uncertainties on $\theta_\mathrm{LD}$ and $\varpi$ in quadrature,
\begin{equation}
    \frac{\sigma_{R_\mathrm{int}}}{R_\mathrm{int}} = \sqrt{ \left(\frac{\sigma_{\theta_\mathrm{LD}}}{\theta_\mathrm{LD}}\right)^{2} + \left(\frac{\sigma_{\varpi}}{\varpi}\right)^{2} } \, .
    \label{eq:Rint_err}
\end{equation}

\begin{table}[]
\renewcommand{\arraystretch}{1.05}
\centering
\caption{The three stars from the catalogue of \citet{Sayeed25} studied in Paper~I with interferometric measurements from \citet{Soubiran24}.}
\begin{threeparttable}
\begin{tabular}{llcc}
\hline
\multicolumn{1}{c}{\thead{\textbf{Name}}} & 
\multicolumn{1}{c}{\thead{\textbf{KIC}}} & 
\multicolumn{1}{c}{\thead{\textbf{parallax} \\ (mas)}} & 
\multicolumn{1}{c}{\thead{$\mathbf{\theta_\mathrm{LD}}$ \\ (mas)}} 
\\
\hline 
Perky & 6106415 & $24.23\pm0.61$ & $0.289\pm0.006$  \\
HD 182736 & 8751420 & $18.34\pm0.58$ & $0.436\pm0.003$  \\
HD 187637 & 6225718 & $18.78\pm0.64$ & $0.231\pm0.006$  \\
\hline 
\end{tabular}
\end{threeparttable}
\vspace*{-3mm}
\label{tab:Inter3adds}
\end{table}
}

\clearpage
\begin{table*}
\renewcommand{\arraystretch}{1.05}
\centering
\caption{Compilation of required data and the associated sources for the interferometric sample.}
\begin{threeparttable}
{\fontsize{8.5}{10}\selectfont
\begin{tabular}{llcccccccc}
\hline
\multicolumn{1}{c}{\thead{\textbf{Name}}} & 
\multicolumn{1}{c}{\thead{\textbf{TIC}}} & 
\multicolumn{1}{c}{\thead{$\mathbf{\nu}_\mathrm{max}$ \\ (\muHz)}} & 
\multicolumn{1}{c}{\thead{\textbf{parallax} \\ (mas)}} & 
\multicolumn{1}{c}{\thead{$\mathbf{\theta_\mathrm{LD}}$ \\ (mas)}} & 
\multicolumn{1}{c}{\thead{\textbf{Source}}} & 
\multicolumn{1}{c}{\thead{$\mathbf{T_\mathrm{eff}}$ \\ (K)}} & 
\multicolumn{1}{c}{\thead{\textbf{Source}}} & 
\multicolumn{1}{c}{\thead{$\mathbf{[\mathrm{Fe}/\mathrm{H}]}$ \\ (dex)}} & 
\multicolumn{1}{c}{\thead{\textbf{Source}}}
\\
\hline 
$\eta$ Boo & 367758676 & $697.9\pm17.8$ & $88.17\pm0.75$ & $2.134\pm0.012$ & (1) & $6128\pm18$ & (1) & $0.25$ & (2) \\
$\zeta$ Her & 43255143 & $718.5\pm24.4$ & $92.63\pm0.60$ & $2.266\pm0.014$ & (1) & $6029\pm19$ & (1) & $0.03$ & (2) \\
$\beta$ Hyi & 267211065 & $1038.1\pm14.9$ & $133.78\pm0.51$ & $2.257\pm0.019$ & (3) & $5872\pm44$ & (3) & $-0.12$ & (2) \\
$\theta$ UMa & 150226696 & $779.3\pm17.8$ & $74.15\pm0.74$ & $1.662\pm0.013$ & (4) & $6256\pm82$ & (4) & $-0.16$ & (4) \\
$\xi$ Gem & 372480991 & $2871.0\pm111.3$ & $57.02\pm0.83$ & $1.401\pm0.009$ & (5) & $6480\pm39$ & (5) & $0.01$ & (5) \\
$\mu$ Her & 460067868 & $1192.5\pm17.5$ & $119.05\pm0.62$ & $1.880\pm0.008$ & (4) & $5425\pm69$ & (4) & $0.23$ & (4) \\
$\eta$ Cas & 445258206 & $2840.0\pm80.7$ & $167.99\pm0.62$ & $1.894\pm0.114$ & (6) & $5907\pm12$ & (7) & $-0.28$ & (7) \\
$\delta$ Eri & 38511251 & $677.6\pm8.3$ & $110.58\pm0.88$ & $2.411\pm0.009$ & (8) & $5027\pm48$ & (8) & $0.07$ & (8) \\
$\delta$ Pav & 409891396 & $2269.8\pm64.4$ & $163.73\pm0.65$ & $1.828\pm0.025$ & (8) & $5604\pm38$ & (8) & $0.33$ & (8) \\
$\beta$ Vir & 366661076 & $1446.3\pm75.9$ & $91.74\pm0.77$ & $1.754\pm0.069$ & (9) & $5456\pm108$ & (9) & $0.12$ & (9) \\
$\gamma$ Lep & 93280676 & $2257.5\pm50.9$ & $111.49\pm0.60$ & $1.871\pm0.112$ & (6) & $6313\pm26$ & (7) & $-0.08$ & (7) \\
$\beta$ Aql & 375621179 & $414.7\pm6.8$ & $72.95\pm0.83$ & $2.133\pm0.012$ & (8) & $5062\pm57$ & (8) & $-0.19$ & (8) \\
$\iota$ Peg & 357336603 & $2101.8\pm126.6$ & $85.06\pm0.71$ & $1.070\pm0.100$ & (6) & $6419\pm141$ & (2) & $-0.11$ & (2) \\
$\gamma$ Ser & 377415363 & $1744.6\pm38.2$ & $89.92\pm0.72$ & $1.161\pm0.055$ & (6) & $6296\pm16$ & (2) & $-0.18$ & (2) \\
$\theta$ Boo & 441709021 & $1354.0\pm108.4$ & $68.63\pm0.56$ & $0.902\pm0.097$ & (9) & $6885\pm370$ & (9) & $-0.02$ & (9) \\
$\iota$ Per & 116988032 & $1855.3\pm33.9$ & $94.93\pm0.67$ & $1.017\pm0.031$ & (10) & $6449\pm117$ & (10)& $0.08$ & (10) \\
$\iota$ Vir & 6029884 & $644.8\pm24.0$ & $46.74\pm0.87$ & $1.222\pm0.061$ & (9) & $6055\pm151$ & (9) & $-0.07$ & (9) \\
$\nu$ And & 189576919 & $1528.0\pm50.6$ & $74.25\pm0.72$ & $1.083\pm0.018$ & (10) & $6114\pm77$ & (10) & $0.08$ & (10) \\
$\theta$ Per & 302158903 & $2314.2\pm166.2$ & $89.03\pm0.79$ & $1.086\pm0.056$ & (6) & $6206\pm23$ & (2) & $0.01$ & (2) \\
$\iota$ Psc & 419919445 & $1416.4\pm53.6$ & $72.51\pm0.88$ & $1.062\pm0.135$ & (6) & $6169\pm56$ & (2) & $-0.14$ & (2) \\
110 Her & 282038438 & $1061.9\pm28.0$ & $52.37\pm0.68$ & $0.905\pm0.022$ & (9) & $6568\pm80$ & (9) & $-0.03$ & (9) \\
$\xi$ Peg & 60716322 & $986.8\pm16.7$ & $61.54\pm0.77$ & $1.091\pm0.008$ & (5) & $6167\pm36$ & (5) & $-0.24$ & (5) \\
$\beta$ CVn & 458445966 & $2385.4\pm136.0$ & $119.46\pm0.83$ & $1.133\pm0.034$ & (4) & $5966\pm117$ & (4) & $-0.19$ & (4) \\
10 Tau & 311092847 & $1284.1\pm63.1$ & $72.89\pm0.78$ & $1.130\pm0.068$ & (6) & $6000\pm59$ & (2) & $-0.08$ & (2) \\
$\lambda$ Ser & 296740796 & $1856.6\pm46.4$ & $82.08\pm0.80$ & $0.982\pm0.056$ & (9) & $6087\pm174$ & (9) & $-0.02$ & (9) \\
$\theta$ Cyg & 27014182 & $1759.1\pm67.1$ & $53.78\pm0.47$ & $0.747\pm0.006$ & (11) & $6853\pm29$ & (11) & $0.06$ & (11) \\
$\gamma$ Dra A & 441804568 & $1232.4\pm19.8$ & $45.38\pm0.82$ & $0.949\pm0.026$ & (5) & $6014\pm90$ & (5) & $-0.17$ & (5) \\
$\chi$ Her & 157364190 & $1045.6\pm17.7$ & $63.08\pm0.54$ & $1.053\pm0.143$ & (9) & $5668\pm386$ & (9) & $-0.45$ & (9) \\
$\lambda$ Aur & 409104974 & $2152.0\pm54.2$ & $79.08\pm0.90$ & $0.940\pm0.056$ & (6) & $5823\pm45$ & (2) & $0.06$ & (2) \\
40 Leo & 95431211 & $1405.8\pm57.9$ & $47.24\pm0.82$ & $0.731\pm0.030$ & (12) & $6447\pm40$ & (12) & $0.09$ & (12) \\
HD5015 & 285544488 & $1399.3\pm52.9$ & $53.85\pm0.60$ & $0.865\pm0.010$ & (5) & $5963\pm44$ & (5) & $0.00$ & (5) \\
36 UMa & 416519065 & $2319.5\pm96.8$ & $77.82\pm0.65$ & $0.778\pm0.014$ & (5) & $6233\pm68$ & (5) & $-0.16$ & (5) \\
70 Vir & 95473936 & $940.6\pm12.9$ & $55.22\pm0.73$ & $0.998\pm0.005$ & (13)& $5494\pm30$ & (13) & $-0.09$ & (13) \\
36 Dra & 233121747 & $1312.0\pm16.9$ & $42.56\pm0.45$ & $0.664\pm0.015$ & (14) & $6638\pm83$ & (14) & $-0.03$ & (14) \\
47 UMa & 21535479 & $2327.9\pm55.1$ & $71.04\pm0.66$ & $0.774\pm0.073$ & (6) & $5862\pm43$ & (15) & $-0.02$ & (15) \\
16 Cep & 366412503 & $643.4\pm11.2$ & $26.67\pm0.45$ & $0.621\pm0.018$ & (14) & $6755\pm106$ & (14) & $-0.20$ & (14) \\
94 Cet & 49845357 & $1267.2\pm99.6$ & $44.69\pm0.75$ & $0.761\pm0.011$ & (14) & $6063\pm45$ & (14) & $0.20$ & (14) \\
HD 33564 & 142103211 & $1736.0\pm61.1$ & $47.66\pm0.52$ & $0.640\pm0.010$ & (16) & $6420\pm50$ & (16) & $0.08$ & (16) \\
$\chi$ Cnc & 302188141 & $1991.8\pm78.8$ & $55.17\pm0.93$ & $0.706\pm0.013$ & (17) & $6130\pm58$ & (17) & $-0.26$ & (17) \\
$\rho$ And & 288294358 & $390.3\pm11.4$ & $20.42\pm0.73$ & $0.754\pm0.014$ & (14) & $5382\pm51$ & (14) & $0.00$ & (14) \\
31 Aql & 359981217 & $1791.8\pm225.2$ & $66.01\pm0.77$ & $0.845\pm0.025$ & (5) & $5787\pm92$ & (5) & $0.33$ & (5) \\
72 Her & 9728611 & $2241.4\pm85.1$ & $69.48\pm0.56$ & $0.725\pm0.012$ & (17) & $5738\pm48$ & (17) & $-0.37$ & (17) \\
$\rho$ CrB & 458494003 & $1664.7\pm99.9$ & $57.38\pm0.71$ & $0.735\pm0.014$ & (16) & $5627\pm54$ & (16) & $-0.22$ & (16) \\
51 Peg & 139298196 & $2485.0\pm97.8$ & $65.10\pm0.76$ & $0.650\pm0.014$ & (14) & $5978\pm67$ & (14) & $0.20$ & (14) \\
15 Peg & 326202925 & $1389.9\pm70.7$ & $36.15\pm0.69$ & $0.542\pm0.005$ & (11) & $6403\pm30$ & (11) & $-0.53$ & (11) \\
HD 195564 & 205591703 & $1167.3\pm46.9$ & $41.26\pm0.87$ & $0.712\pm0.031$ & (17) & $5421\pm118$ & (17) & $0.06$ & (17) \\
HD 190360 & 105999792 & $2358.2\pm74.1$ & $62.92\pm0.62$ & $0.663\pm0.005$ & (11) & $5557\pm22$ & (11) & $0.17$ & (11) \\
HD 89744 & 8154501 & $1028.4\pm108.5$ & $25.65\pm0.70$ & $0.556\pm0.032$ & (18) & $5927\pm185$ & (18) & $0.21$ & (18) \\
HD 49933 & 281812116 & $1997.7\pm117.8$ & $33.45\pm0.84$ & $0.445\pm0.012$ & (19) & $6640\pm100$ & (19) & $-0.38$ & (19) \\
HD 38529 & 200093173 & $622.5\pm27.9$ & $23.57\pm0.92$ & $0.573\pm0.049$ & (20) & $5634\pm21$ & (21) & $0.38$ & (21) \\
\hline
\multicolumn{10}{c}{\large\thead{Additions from \citet{Soubiran24}}} \\
\hline
HD 181096 & 158985758 & $960.0\pm15.0$ & $23.57\pm0.57$ & $0.443\pm0.007$ & (2) & $6390\pm51$ & (2) & $-0.25$ & (2) \\
HD 167042 & 353988778 & $249.2\pm12.5$ & $20.00\pm0.51$ & $0.831\pm0.007$ & (2) & $4987\pm66$ & (2) & $0.04$ & (2) \\
HD 90043 & 1713457 & $203.0\pm10.0$ & $13.37\pm0.87$ & $0.659\pm0.009$ & (2) & $4801\pm83$ & (2) & $-0.03$ & (2) \\
HD 198149 & 372682437 & $228.0\pm2.0$ & $69.73\pm0.49$ & $2.882\pm0.088$ & (2) & $4751\pm73$ & (2) & $-0.13$ & (2) \\
\hline
\end{tabular}
}
\end{threeparttable}
\tablefoot{The \numax values are taken from Table~D.4 of \citet{Lund25} for the majority of stars and from \citet{Soubiran24} for the additions. All parallaxes are from Hipparcos \citep{vanLeeuwen07}.}
\tablebib{
(1)~\citet{Baines14};
(2)~\citet{Soubiran24};
(3)~\citet{North07};
(4)~\citet{Baines18};
(5)~\citet{Boyajian12a};
(6)~\citet{Mennesson14};
(7)~\citet{Soubiran22};
(8)~\citet{Rains20};
(9)~\citet{Baines23};
(10)~\citet{Baines21};
(11)~\citet{Karovicova22};
(12)~\citet{Maestro13};
(13)~\citet{Kane15};
(14)~\citet{Ligi16};
(15)~\citet{Sharma24};
(16)~\citet{vonBraun14};
(17)~\citet{Boyajian13};
(18)~\citet{Schaefer18};
(19)~\citet{Bigot11};
(20)~\citet{Baines08};
(21)~\citet{Sousa24}
}
\vspace*{-3mm}
\label{tab:InterSamp}
\end{table*}

\clearpage
\section{MCMC Implementation and Convergence Diagnostics}\label{app:pystan}
We implemented the hierarchical model using Stan via the \textsc{CmdStan} interface. Sampling was performed with the No-U-Turn Sampler (NUTS; \citealt{Hoffman14}), an adaptive variant of Hamiltonian Monte Carlo \citep{Betancourt13}. We ran $N_\mathrm{chains} = 4$ independent chains, each for a total of $10000$ iterations. The first third of each chain was discarded as warm-up, leaving $4 \times 6666 = 26\,664$ posterior draws retained for inference. The step-size adaptation targets an acceptance rate of $\delta = 0.95$, and the maximum tree depth was set to 12, both settings reflecting stricter sampling criteria than default.

Chains were initialised at the observed values for all latent stellar parameters ($\log A_\mathrm{gran}$, $\log b$, $\log T_\mathrm{eff}$, $\log R$, $[\mathrm{Fe/H}]$), with the regression coefficients set to representative starting points ($x = 0.5$, $y = -0.7$, $z = 1.5$, $w = 0.1$, $C = 0.01$) and intrinsic scatter at $\sigma = 0.03$. A small random jitter ($5\%$ Gaussian noise) was applied independently to global parameters across chains to enable independent posterior exploration.

Convergence is assessed through three complementary criteria applied to all regression parameters ($C$, $x$, $y$, $z$, $w$, $\sigma$):
\begin{enumerate}
    \item \textbf{Gelman-Rubin statistic} $\left(\hat{R}\right)$. We required $\hat{R} < 1.025$ for every parameter, confirming that all chains had converged to the same target distribution \citep{Gelman92}.
    \item \textbf{Effective sample size} (ESS). We required $\mathrm{ESS}/N_\mathrm{total} > 0.001$, where $N_\mathrm{total}$ is the total number of post-warmup draws. Values below this threshold indicate excessive autocorrelation and insufficient posterior exploration.
    \item \textbf{Energy Bayesian Fraction of Missing Information} (E-BFMI). Computed per chain as the ratio of the variance in successive energy differences to the marginal energy variance, E-BFMI~$< 0.2$ signals inadequate momentum resampling and potential pathologies in the posterior geometry \citep{Betancourt16}. We required E-BFMI~$\geq 0.2$ for all chains.
\end{enumerate}
In addition, the built-in \texttt{diagnose()} routine from \textsc{CmdStan} was used to flag any divergent transitions or saturated treedepth after every fit. Posterior samples and diagnostics were saved in \textsc{ArviZ} NetCDF format \citep{Kumar19}, and trace plots recorded to be inspected visually if necessary.

The hierarchical model may be seen below, written in Stan following the setup outlined throughout Sect.~\ref{sec:Methods}.

\begin{lstlisting}
data {
    int<lower=0> N;
    vector[N] log_amp_obs;
    vector[N] log_charb_obs;
    array[N] matrix[2, 2] gran_cov;
    vector[N] log_teff_obs;
    vector<lower=0>[N] log_teff_err;
    vector[N] log_R_obs;
    vector<lower=0>[N] log_R_err;
    vector[N] feh_obs;
    array[N] int<lower=0, upper=1> has_feh_obs;
}
parameters {
    real C;
    real x;
    real y;
    real z;
    real w;
    vector[N] log_teff_true;
    vector[N] log_R_true;
    vector[N] log_amp_true;
    vector[N] log_charb_true;
    vector<lower=-4, upper=1>[N] feh_prior;
    real<lower=0> sigma;
}
model {
    vector[N] log_R_pred;
    vector[N] feh_effective;

    for (i in 1:N) {
        if (has_feh_obs[i] == 1) {
            feh_effective[i] = feh_obs[i];
        } else {
            feh_effective[i] = feh_prior[i];
            feh_prior[i] ~ normal(-0.10, 0.4);
        }
    }

    log_R_pred = C + x*log_amp_true + y*log_charb_true
                    + z*log_teff_true + w*feh_effective;

    log_R_pred   ~ normal(log_R_true,    sigma);
    log_R_obs    ~ normal(log_R_true,    log_R_err);
    log_teff_obs ~ normal(log_teff_true, log_teff_err);

    for (i in 1:N) {
        vector[2] obs;
        vector[2] true_vals;
        matrix[2, 2] L;
        obs[1] = log_amp_obs[i];       
        obs[2] = log_charb_obs[i];
        true_vals[1] = log_amp_true[i]; 
        true_vals[2] = log_charb_true[i];
        L = cholesky_decompose(gran_cov[i]);
        obs ~ multi_normal_cholesky(true_vals, L);
    }

    C   ~ normal(0, 2);
    x     ~ normal(0, 0.5);
    y     ~ normal(0, 0.5);
    z     ~ normal(0, 0.5);
    w    ~ normal(0, 0.5);
    sigma ~ normal(0, 0.2);

    log_teff_true  ~ normal(log_teff_obs,  0.5);
    log_R_true     ~ normal(log_R_obs,     0.5);
    log_amp_true   ~ normal(log_amp_obs,   0.5);
    log_charb_true ~ normal(log_charb_obs, 0.5);
}
\end{lstlisting}

\section{Derivation results for models J and T}\label{app:derivres}
This appendix presents the derivation results of background models J and T, analogous to Fig.~\ref{fig:DerivH} for model H. The values obtained for the global regression parameters seen in the cornerplots are compiled in Table~\ref{tab:GlobParams}. 
\begin{figure*}[]
    \includegraphics[width=\linewidth]{Figures/DerivJ.jpg}
    \centering
    \vspace*{-5.5mm}
    \caption{Results of the granulation-radius relation derivation for the hybrid background model (J). \textit{Left panel:} A cornerplot of the inferred global regression parameters of Eq.~\ref{eq:radscal}. The summary statistics are given as the median and 16/84 percentiles of the posteriors. \textit{Right panel:} The granulation-inferred stellar radii plotted against the asteroseismic radii for each star. In the main panel we use the inferred \teff for the prediction, while the insert shows the results obtained when using the observed \teff. The obtained fractional root-mean-square scatter, $\Delta R/R$, are shown for each. The residuals are shown below, coloured according to the granulation background model preference as indicated by the legend.}
    \label{fig:DerivJ}
    \vspace*{-2mm}
\end{figure*}
\begin{figure*}[]
    \includegraphics[width=\linewidth]{Figures/DerivT.jpg}
    \centering
    \vspace*{-5.5mm}
    \caption{Results of the granulation-radius relation derivation for the three-component background model (T). \textit{Left panel:} A cornerplot of the inferred global regression parameters of Eq.~\ref{eq:radscal}. The summary statistics are given as the median and 16/84 percentiles of the posteriors. \textit{Right panel:} The granulation-inferred stellar radii plotted against the asteroseismic radii for each star. In the main panel we use the inferred \teff for the prediction, while the insert shows the results obtained when using the observed \teff. The obtained fractional root-mean-square scatter, $\Delta R/R$, are shown for each. The residuals are shown below, coloured according to the granulation background model preference as indicated by the legend.}
    \label{fig:DerivT}
    \vspace*{-2mm}
\end{figure*}

\newpage
\section{Origin of predictive power}\label{app:PredConstraints}
\begin{figure*}[]
    \centering
    \includegraphics[width=18cm]{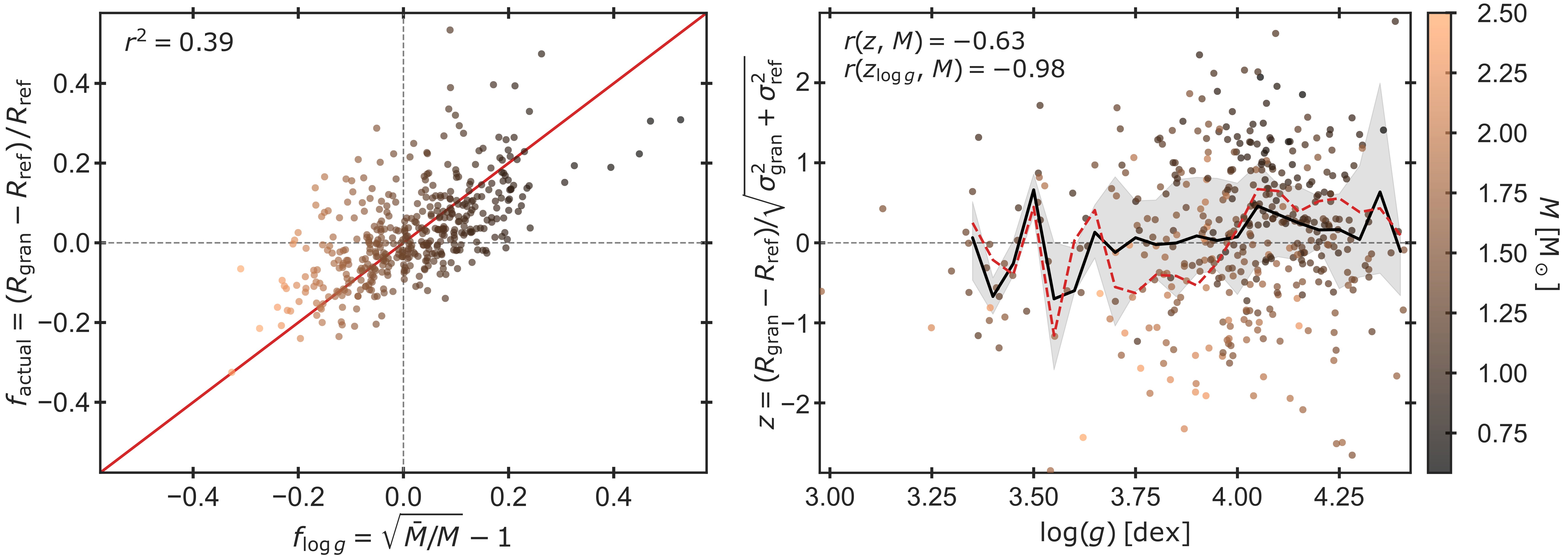}
    \vspace*{-1mm}
    \caption{Obtained residuals compared to \logg-predicted residuals for the 428 stars in the validation sample with seismic masses, which are indicated by the colour bar. \textit{Left panel:} The fractional radius errors obtained compared to the prediction by a pure \logg-inverter (red line). \textit{Right panel:} The standardised residuals of our granulation-inferred radii plotted against \logg. For thin 0.05 dex bins, we show the median z-score by the black line and the \logg-prediction with the dashed red line. The inter-quartile range within each bin is shown by the grey shaded region.}
    \label{fig:logg_rad_const}
    \vspace*{-3mm}
\end{figure*}
Granulation signals are known to constrain surface gravity \citep{Bastien16, Bugnet18, Pande18}. In this work, however, we constructed a hierarchical model that directly infers stellar radius. This raised a key question: does the model extract radius-sensitive information from the granulation signal, or does it effectively infer \logg and convert it to a radius via $g = GM/R^2$, implicitly adopting the mass distribution of the calibration sample?

Under the null hypothesis that the granulation observables constrain only \logg, the inferred radius is set by the characteristic mass of the calibration sample, which we summarise by its mean $\tilde{M} = 1.36\,M_\odot$. For a star with true mass $M \neq \tilde{M}$, this leads to a predictable bias. In this case, the inferred radius becomes $R_\mathrm{gran} = R_\mathrm{ref}\sqrt{\tilde{M}/M}$ and the resulting fractional error is
\begin{equation}\label{eq:flogg}
    f_{\logg} = \frac{R_\mathrm{gran} - R_\mathrm{ref}}{R_\mathrm{ref}} = \sqrt{\frac{\tilde{M}}{M}} - 1 \ .
\end{equation}
This expression represents the limiting case of a pure \logg-inverter and therefore provides a conservative upper bound on mass-driven systematics. It predicts systematic overestimation of the resulting radii for $M < \tilde{M}$ and underestimation for $M > \tilde{M}$. The left panel of Fig.~\ref{fig:logg_rad_const} compares the observed fractional radius errors, $f_\mathrm{actual}$, with $f_{\logg}$. A pure \logg-inverter would place the data along the one-to-one relation, whereas genuine radius sensitivity would produce no correlation, with points clustering around zero. The observed distribution lies between these limits: the data show a clear but incomplete alignment with the predicted trend. This indicates that the model is influenced by \logg, but also captures additional information that constrains the radius directly. It was fully expected that some constraining power would come through the \logg channel. Yet, Fig.~\ref{fig:logg_rad_const} shows that our hierarchical model is not just a \logg-inverter, but captures additional information through out joint use of the granulation observables $A_\mathrm{gran}$ and $b$. The residual mass trend is, however, real, manifesting as predictable over/under-estimation for stars below/above the mean mass $\tilde{M}$ of the calibration sample. This is a limitation of calibrating on a sample of solar-like oscillators that spans a relatively narrow mass range \citep{Serenelli17}, compounded by the need to exclude stars cooler than the Sun ($\numax> 3000$ \muHz).

To quantify these findings, we compute the Pearson correlation coefficient\footnote{See e.g. https://docs.scipy.org/doc/scipy/reference/generated/scipy.stats.pearsonr.html} between $f_\mathrm{actual}$ and $f_{\logg}$. The resulting measure $r^2$ is the fraction of the observed variance in $f_\mathrm{actual}$ that co-varies linearly with $f_{\logg}$: a value near 1 would indicate a pure \logg-inverter; a value near 0 indicates no coupling. Calculated for the validation sample we find $r^2=0.39$ (i.e. $r = 0.62$). Hence, at most $39\%$ of the variance can be attributed to the \logg channel, with the remaining $61\%$ being orthogonal to it. This latter fraction absorbs observational noise from the granulation fit and \teff, intrinsic scatter in the derived granulation-radius relations, and crucially, genuine radius-specific information carried by the granulation observables which is encoded into the hierarchical model.

The test above was global, assessing whether a star’s fractional radius error followed the $\logg$-based prediction across the full sample. As a complementary approach, we fixed the evolutionary stage and examined whether stellar mass continued to drive the z-score residuals within each slice. We binned the validation stars in narrow $\logg$ intervals of 0.05 dex, where $\logg$ was computed from the seismic mass and reference radius from \citet{Sayeed25}. A pure $\logg$-inverter is unable to resolve the individual stellar masses and will therefore assign identical radii within a fixed $\logg$ bin, implying $R \propto \sqrt{M}$. For each star, we computed the corresponding expected z-score as
\begin{equation}\label{eq:zlogg}
    z_{\logg}=f_{\logg}\frac{R_\mathrm{ref}}{\sigma_\mathrm{tot}}, \quad \sigma_\mathrm{tot}=\sqrt{\sigma_\mathrm{gran}^2 + \sigma_\mathrm{ref}^2} \ .
\end{equation}
The bin medians of $z_{\logg}$ and the actual z-scores from Sect.~\ref{subsec:AppValid} are shown in Fig.~\ref{fig:logg_rad_const}. While the two trends co-vary, which is expected as $z_{\logg}$ was computed from the same stellar masses that drive the residuals, their amplitudes differ significantly. The actual z-scores generally show a flatter trend across evolutionary stages, particularly along the upper MS and early SGB where the sample is densest. The interquartile ranges further show that the scatter within each bin is dominated by factors other than mass. This difference may again be expressed quantitatively by the Pearson correlation coefficient. For the actual z-scores, we find $r(z_\mathrm{score}, M) = -0.626$, whereas the theoretical \logg expectation yields $r(z_{\logg}, M) = -0.978$, close to $-1$ by construction since $z_{\logg}$ is a monotone function of $M$. The ratio of these values, $\sim 0.64$, shows that the amplitude of the observed mass dependence is only about two-thirds of that expected from a pure \logg-inverter. This is consistent with the $r=0.62$ found above.

Taken together, these two tests lead to a consistent conclusion: while the inferred radii retain a measurable imprint of \logg, this dependence is significantly diluted. The granulation observables, when combined within the hierarchical framework, carry independent information that constrains stellar radius beyond surface gravity alone.

\end{appendix}
\end{document}